\documentclass[aps,prd,final,amsmath,amssymb,superscriptaddress,floatfix,showpacs,footinbib]{revtex4}
\usepackage{float,graphicx}

\newcommand{\dd}{{\rm d}}

\newcommand{\GeV}{\,\mathrm{GeV}}
\newcommand{\eV}{\,\mathrm{eV}}
\newcommand{\meV}{\,\mathrm{meV}}
\newcommand{\cm}{\,\mathrm{cm}}

\begin{document}

\title{Testing Chameleon Theories with Light Propagating through a Magnetic Field}

\author{Philippe~Brax}
\affiliation{ Service de Physique
Th\'eorique CEA/DSM/SPhT, Unit\'e de recherche
associ\'ee au CNRS, CEA-Saclay F-91191 Gif/Yvette
cedex, France.}
\author{Carsten van de Bruck}
\affiliation{ Department of Applied Mathematics, The University of Sheffield,
 Hounsfield Road, Sheffield S3 7RH, United Kingdom}
\author{Anne-Christine Davis}
\affiliation{Department of Applied Mathematics and Theoretical Physics,  Centre for Mathematical Sciences,  Cambridge CB2 0WA, United Kingdom}
\author{David F. Mota}
\affiliation{Institut f\"ur Theoretische Physik,
Universit\"at Heidelberg, Philosophenweg 16/19,  D-69120 Heidelberg, Germany}
\author{Douglas Shaw}
\affiliation{ Centre for Mathematical Sciences, Cambridge CB2 0WA, United Kingdom}

\begin{abstract}
It was recently argued that the observed PVLAS anomaly can be
explained by chameleon field theories in which large deviations
from Newton's law can be avoided. Here we present the predictions
for the dichroism and the birefringence induced in the vacuum by a
magnetic field in these models. We show that chameleon particles
behave very differently from standard axion-like particles (ALPs).
We find that, unlike ALPs, the chameleon particles are confined
within the experimental set-up. As a consequence,  the
birefringence  is always bigger than the dichroism in PVLAS-type
experiments.
\end{abstract}


\maketitle

\section{Introduction}
Light scalar fields are common in theories of physics beyond the standard
model. These scalar fields can couple to both the standard model
fields as well as to new types of matter. Since experiments have
not yet detected new forces, it means that the force mediated by
these scalar fields is either very weak (with a strength less than
gravity) or short-ranged in the laboratory. In string theory, for
example, there are many moduli fields which couple to matter with
gravitational strength. The chameleon mechanism provides a way to
suppress the forces mediated by these scalar fields via non-linear
field self-interactions and interactions with the ambient matter
\cite{chamKA,chamstrong}. As a result, the masses of the scalar
fields become dependent on the ambient matter density; this is the
reason why these fields have been dubbed chameleon fields. If the
observed accelerated expansion of the universe is due to a
chameleon-like field, it has interesting cosmological consequences
\cite{chamcos,chamstruc}.

An obvious way to look for chameleon fields are gravitational
experiments in different environments \cite{chamKA,chamstrong}.
Additionally, it was recently pointed out that chameleon fields
are a natural way to reconcile the PVLAS and CAST results
\cite{chamPVLAS}. In 2006, it was reported that the PVLAS
experiment had detected light polarization rotation in the vacuum
in the presence of a magnetic field, \cite{PVLAS}. Recently there
has been a lot of activity concerning the theoretical
interpretation of this detection, see e.g.
\cite{ringwald1}--\cite{ahlers2} for recent work. In particular,
this finding could be seen as evidence for the presence of an
axion-like particle (ALP) or milli-charged particles.  The PVLAS
2006 results can be explained if the mass of the ALP is of order
$m_{\rm ALP} \approx 10^{-3}$~eV and the inverse coupling constant
to two photons is $M \approx 10^5$~GeV. These results are in
direct conflict with the CAST results, because, for these
parameters, axions emitted in the sun should have been detected by
this experiment. However, as was pointed out in \cite{chamPVLAS},
assuming the values for $m$ and $M$ given above, both results can
be explained if the particle behaves like a chameleon field,
because the mass of the field inside the PVLAS experiment would
then be very different from the mass of the field inside the Sun.
In the Sun, the chameleon mass is so high that chameleon particles
cannot be photo-produced. Additionally, it was shown in \cite
{chamPVLAS} that the new force mediated by the chameleon field is
not in conflict with current experiments.

Although in a context different from chameleon theories, the
potential resolution of conflict between the CAST and the PVLAS
experiments provided by particles with an environmentally dependent
mass was also discussed in Ref. \cite{jaeckel}. It was also pointed
out that if the particle mass outside the cavity was much larger
than inside, then ALPs would be reflected in the same way as the
photons and \emph{not} escape from the Fabry-Perot cavity. As an
implication, there would be no signal found in "light shining
through the wall" experiments \cite{jaeckel}. As the BMV collaboration seems to have observed \cite{nolight}.  It should be noted
that in Ref. \cite{jaeckel} the ALPs were assumed to reflect in
exactly the same way as the photons;  we see below that this is not
the case if the ALP is a chameleon. The predictions of the chameleon
model are therefore very different from the ones of reflecting ALP
model consider in Ref. \cite{jaeckel}.

Recently new PVLAS results have been reported \cite{PVLAS07}.
These new results do \emph{not} confirm the previously reported
rotation. Indeed they find no evidence for either rotation or
ellipticity with a $2.3$~T magnetic field.  The initial PVLAS results
were found using a $5.5$~T magnetic field. With a $5.5$~T field,
the new results additionally show no evidence for any rotation,
however a non-zero ellipticity is still detected. In the context
of standard ALPs (but not necessarily chameleon fields), however,
the detected ellipticity is excluded with a $99\%$ confidence by
the null result found with the $2.3$~T magnetic field. Although it
might be an artefact, the ellipticity for $B=5.5$~T can be
explained with a chameleon model with $m\approx 10^{-3}$~eV
and $M\approx 10^6$~GeV. Future experiments will give us decisive
clues later this year.

In this paper we study the behaviour of chameleon particles inside
optical experiments similar to PVLAS. The aim of this paper is to
derive the expression for the rotation and the ellipticity of the
laser light. As we will show, chameleon particles and standard
ALPs behave very differently. Firstly, because the mass of the
particles depend on the environment and they do not have enough
energy, chameleon particles cannot leave the experimental region.
In contrast to ALPs, which do leave the interaction region,
chameleon particles are therefore trapped inside the experimental
set-up. Secondly, because the mass of the particles varies inside
the experiment, chameleon particles and photons reflect
differently. As we will discuss, this results in a very different
form for the expressions for the predicted rotation and
ellipticity.

The paper is organized as follows: In Section \ref{sec:chammodel}
we present the chameleon model in more detail. In Section
\ref{sec:cylinder} we discuss how the chameleon mass behaves
inside the experimental set-up. The behaviour of chameleon
particles in experiments like PVLAS is described in Section
\ref{sec:prop}. The predictions made by the chameleon model are
discussed in Section  \ref{sec:predictions}. We conclude in
Section \ref{sec:con}. Mathematical details can be found in the
Appendices.

\section{The chameleon model}\label{sec:chammodel}
Chameleon theories are essentially scalar field theories with a self-interaction
potential and couplings to matter; they are specified by the action
\begin{eqnarray}
S&=&\int d^4x \sqrt{-g}\left(\frac{1}{2\kappa_4^2}R-
g^{\mu\nu}\partial_\mu\phi \partial_\nu \phi -V(\phi)
-\frac{e^{\phi/M}}{4} F^2\right)\nonumber \\
&+& \sum S_m^{(i)}( e^{\phi/M_i}
g_{\mu\nu},\psi_m^{(i)}) \label{action}
\end{eqnarray}
where the $S_m^{(i)}$ and $\psi_{m}^{(i)}$ are respectively the matter actions and matter fields. The couplings to matter and the electromagnetic sector are specified by the mass scales $M_i$ and $M$ respectively. For simplicity we take the
coupling of the scalar field to matter and to the electromagnetism to be universal (i.e. $M_i = M$).
In more general theories one would not expect a truly universal coupling. Our conclusions are, however, not affected by this assumption. The coupling to matter implies
that particle masses in the Einstein frame depend on the value of $\phi$
\begin{equation}
m(\phi)= e^{\phi/M} m_0
\end{equation}
where $m_0$ is the bare mass as appearing in $S_m$. The strength of the chameleon to matter coupling is given by
\begin{equation}
\beta= \frac{M_{\rm Pl}}{M}.
\end{equation}
where $M_{Pl} = 1/\sqrt{8\pi G} \approx 2.4 \times 10^{18}\GeV$.
As we show in Section \ref{sec:prop}, if a theory such as this is
to be detected by axion searches one must require that $M \ll
M_{Pl}$.  This
implies that on scales smaller than $\hbar c / m_{\phi}$, where
$m_{\phi}$ is the mass of $\phi$, the chameleon force between
matter particles is $2(M_{Pl}/M)^2 \gg 1$ times stronger than
their mutual gravitational attraction.  If the mass, $m_{\phi}$,
of $\phi$ is a constant then one would then have to require that
$m_{\phi} \gg 1\meV$ otherwise such a theory would already be
ruled out by experimental tests of Newton's law.  Another, and
potentially far more interesting prospect, is that the mass of the
scalar field grows with the background density of matter.  In high
density regions it could then be large enough to satisfy the
constraints coming from tests of gravity, whilst being small enough
to produce detectable alterations to the standard physical laws in
low density regions. Scalar fields that have this property are
said to be Chameleon fields. In addition to the couplings to
matter, chameleon fields have non-linear self-interactions described by a
potential $V(\phi)$. Assuming an exponential coupling to matter of
the form given by Eq. (\ref{action}), a scalar field theory will
have a chameleon mechanism, for some range of $\phi$, provided
that:
\begin{equation}
V^{\prime}(\phi) < 0, \quad V^{\prime \prime} > 0,\quad V^{\prime \prime \prime}(\phi) < 0, \label{chamcond}
\end{equation}
where $V^{\prime} = \dd V / \dd\phi$.  Whether or not the chameleon mechanism is strong enough
to evade current experimental constraints depends partially on the details of the theory, i.e. $V(\phi)$ and $M$,
 and partially on the initial conditions \cite{chamcos}.  For
exponential couplings and a potential of the form
\begin{equation}\label{poti}
V(\phi)= \Lambda^4\exp (\Lambda^n/\phi^n) \approx \Lambda^4 +
\frac{\Lambda^{4+n}}{\phi^n}
\end{equation}
the chameleon mechanism can in principle hide the field such that
there is no conflict with current laboratory, solar system or
cosmological experiments \cite{chamKA,chamcos}. Importantly, the
chameleon mechanism is strong enough in such theories to allow strongly
coupled theories with $M \ll M_{Pl}$ to have remained thus far
undetected \cite{chamstrong}.

The first term in the potential, $V$, corresponds to an effective
cosmological constant whilst the second term is a Ratra-Peebles
inverse power law potential. If one assumes that $\phi$ is also
responsible for late-time acceleration of the universe then one
must require $\Lambda\approx 2.3 \times 10^{-12}\GeV$.

The evolution of the chameleon field in the presence of both matter and an external
magnetic field is determined by the effective potential:
\begin{equation}
V_{\rm eff}(\phi)=V(\phi) + \rho_{\rm eff} e^{\phi/M}
\end{equation}
where
\begin{equation}
\rho_{\rm eff} = \rho_{\rm matter} + \frac{B^2}{2}
\end{equation}
and $B$  is the magnetic field. As a result, even though $V$ has a runaway form,  the effective
potential has a minimum at $\phi = \phi_{\rm min}(\rho_{\rm eff})$ where
\begin{equation}
V^{\prime}_{\rm eff}(\phi_{\rm min}) = 0 = V^{\prime}(\phi_{\rm min}) + \frac{1}{M}\left(\rho_{\rm matter} + \frac{B^2}{2}\right).
\end{equation}
In the presence of ambient matter and a magnetic field, the field evolves towards this minimum. The mass of small perturbations in $\phi$ about $\phi_{\rm min}$ is given by
\begin{eqnarray}\label{mas1}
m_{\phi}(\phi_{\rm min}) = \left(n(n+1)\frac{\Lambda^{n+4}}{\phi^{n+2}_{\rm min}}\right)^{1/2},
\end{eqnarray}
and we have that
\begin{eqnarray}\label{mas2}
\phi_{\rm min} = \left(\frac{2n\Lambda^{4+n}M}{2\rho_{\rm matter}+B^2}  \right)^{1/(1+n)}.
\end{eqnarray}
In some circumstances, however, $\phi$ is unable to change quickly enough to actually reach
$\phi_{\rm min}$. In particular, this behaviour can occur inside a low density cavity. If the radius, $R$, of
the cavity is too small, then $\phi$ does not reach its effective minimum and instead, as we
show in Section \ref{sec:cylinder}, $m_{\phi} \sim 2/R$ for $R \ll 2/m_{\phi}(\phi_{\rm min})$. Since laboratory
searches for vacuum magnetic dichroism and birefringence generally employ such a cavity, one must be particularly
wary of this behaviour when making predictions for what such experiments should detect. In particular
the dependence of $m_{\phi}$ on $B$ and $\rho_{\rm matter}$ inside the experiment depends on the
details of the set-up.

The chameleon mass, $m_{\phi}$, depends therefore on a number of factors. In searches
for dichroism and birefringence in a vacuum, the mass depends on
the details of the experimental set-up itself: the size of the
cavity, the magnetic field $B$ and the density of matter in the
laboratory vacuum. The mass is \emph{not} a fundamental parameter;
the fundamental parameters of our model are $\Lambda$, $M$ and $n$.
It is also important to note that, not only is $m_{\phi}$ not a
fundamental parameter, but that it is generally very different in
different parts of  experiments.   We discuss the behaviour of
$m_{\phi}$ in laboratory searches for axion-like-particles (ALPs)
in the next section.

\section{The chameleon mass in the laboratory}\label{sec:cylinder}
In laboratory searches, such as the PVLAS \cite{PVLAS}, and Q$\&$A
\cite{QA} experiments, for the dichroism and birefringence induced
by a magnetic field, light propagates in a Fabry-Perot cavity  with
radius $R$.  The interaction region, i.e. the region where the
magnetic field is turned on ($B \neq 0$), has length $L$. To
increase the strength of any signal the light is reflected $N$
times, and the mirrors are located a distance $d$ from either end of
the interaction region.  We label the density of the vacuum matter
inside the cavity by $\rho_{\rm gas}$.    For example in the  PVLAS
experiment $L = 100\cm$, $d = 270\cm$, $R = 12.5\,{\rm mm}$ and
$\rho_{\rm gas} \approx 2 \times 10^{-4}\,{\rm g\,cm}^{-3}$. Before
we can make predictions for the dichroism and birefringence we need
to know the value of $m_{\phi}$ along the path of the photon.
Although we are primarily concerned with power-law type potentials,
in most of the discussion  in this section we do not assume any
particular form of $V$, we only require that it satisfies the
chameleon field theory conditions given by Eq. (\ref{chamcond}).

The cavity is a cylinder with radius $R$.  Outside of the cavity
(i.e. in the walls of the cavity and the surrounding magnet),
$\phi$ must lie close to its effective minimum, which we label by $\phi
= \phi_{\infty}$.  If this were not the case then the walls of two
such cavities would feel a force that would be $2(M_{Pl}/M)^2 \gg
1$ times their gravitational attraction.  Such a force is easily
ruled out by experimental tests of gravity.   We define $r$ to be
the radial distance from the centre of the cavity and define
$\phi_{0} = \phi(r=0)$.  We approximate the potential inside and
outside the cavity as a quadratic function. Outside the cavity,
the expansion is around the minimum where
$V'_{\rm eff}(\phi_{\infty})=0$. Inside we expand around
$\phi_0$ which is left unknown. This leads to two equations for
the field $\phi$. For $r\ge R$ we have
\begin{equation}
\phi'' +\frac{1}{r}\phi' -m_\infty^2 (\phi-\phi_\infty)=0,
\end{equation}
whereas for $r\leq R$ the equation reads
\begin{equation}
\phi'' +\frac{1}{r} \phi' -m_0^2 (\phi-\phi_0)=V^{\prime}_{{\rm eff}\,0},
\end{equation}
where $V^{\prime}_{{\rm eff}\,0} = V^{\prime}(\phi_{0}) + \rho_{\rm eff}e^{\phi_{0}/M}$. Note that
in most cases $\phi_0$ is not $\phi_c$ where $V_{\rm eff}^{\prime}(\phi_c)=0$. We define
 $V_{c}^{\prime} = V^{\prime}(\phi_c)$ and $V^{\prime}_{0} = V^{\prime}(\phi_0)$,
 we can then write $V^{\prime}_{\rm eff,0} = V^{\prime}_{0}-V^{\prime}_{c}$.

The non-singular solution for $r\leq R$ is a combination of the Bessel functions
$J_0$ and $N_0$. However, since $N_0$ diverges logarithmically at the origin, we ignore
the term containing $N_0$ and therefore the solution inside the cavity reads
\begin{equation}
\phi= CJ_0(im_0r) +\phi_0 -\frac{V_0^{\prime}-V_{c}^{\prime}}{m_0^2}.
\end{equation}
On the other hand, for $r\ge R$, the solution is
\begin{equation}
\phi=A( J_0(im_\infty r) -i N_0(im_\infty r)) +\phi_\infty
\end{equation}
Matching both solutions and their first derivatives at $r=R$ gives the following
conditions for $A$ and $C$:
$$
A=\frac{m_0 J'_0(im_0R)(\phi_\infty -\phi_0
+(V_0^{\prime}-V_c^{\prime})/m_0^2)}{m_\infty J_0(im_0R)
(J'_0-iN_0')(im_\infty R) -m_0 J'_0(im_\infty
R)(J_0-iN_0)(im_\infty R)},
$$
and
$$
C= \frac{m_\infty}{m_0}\frac{(J'_0-iN_0')(im_\infty R)}{J'_0(im_0R)}A.
$$
Now, since $\phi$ must lie very close the effective minimum inside the cavity walls, we must have
$$
m_\infty R  \gg 1.
$$
The solution inside the cavity then simplifies drastically
\begin{equation}
\phi= \frac{\phi_\infty-\phi_0+\frac{V_0^{\prime}-V_c^{\prime}}{m_0^2}}{J_0(im_0R)}J_0(im_0r)
+ \phi_0 -\frac{V_0^{\prime}-V_c^{\prime}}{m_0^2}.
\end{equation}
Evaluating this equation at $r=0$ and imposing $\phi(r=0) = \phi_0$ leads to
\begin{equation}
\phi_\infty -\phi_0= \frac{V_0^{\prime}-V_c^{\prime}}{m_0^2} \left(J_0(im_0R) - 1 \right) \label{m0Reqn}
\end{equation}
There are now two relevant situations:

\subsection{$m_c R \gg 1$}
Since $m_{0} \le m_{c}$, $m_0 R\gg 1$ in this case.  It follows from Eq. (\ref{m0Reqn}) that $V^{\prime}_0 \approx V_{c}^{\prime}$ and therefore that
$m_0 \approx m_c$.

\subsection{$m_c R \ll 1$}
If $V^{\prime}_{0}/V^{\prime}_{c} - 1 \lesssim {\cal O}(1)$ that would
require $m_{0} \approx m_{c}$ and by the matching condition $m_{0}R \approx m_{c}R \gtrsim {\cal O}(1)$.
If $m_c R \ll 1$ we must therefore have $V^{\prime}_{0}/V^{\prime}_{c} - 1 \gg 1$, and so
$$
1+ J_{0}(im_{0}R) = \frac{m_{0}^2 (\phi_0 - \phi_{\infty})}{\vert V^{\prime}_{0} \vert}.
$$
The right hand side of this equation is generally ${\cal O}(1)$ or
smaller and so $m_{0}R = {\cal O}(1)$ which implies $m_{0} \ll
m_{\infty}$ and $\phi_{\infty} \ll \phi_{0}$ due to the runaway
form of the potential.  For the inverse power-law potential
$V_0'/m_0^2= -\phi_0/(n+1)$, and the matching equation therefore gives
\begin{equation}\label{relation1}
J_0(im_0 R) = n + 2,
\end{equation}
which implies, as expected, that $m_0R ={\cal O}(1)$ for $n = {\cal O}(1)$. Expanding the
Bessel function for small $m_{0}R \ll 4$ in the last equation gives
\begin{equation}\label{relation2}
m_0R \approx 2\sqrt{n+1}.
\end{equation}
A slightly better approximation for ${\cal O}(1)$ values of $n$ is
given by:
\begin{equation}\label{relation3}
m_{0}R \approx 2\sqrt{2}(\sqrt{n+2}-1)^{1/2}.
\end{equation}
Note that if $m_{c}R \sim {\cal O}(1)$ then we could not ignore
the $V_{c}^{\prime}$ term in the matching condition, but since
$V_{c}^{\prime} < 0$ we would have:
$$
J(im_{0}R) - 1 >  n+2,
$$
and so $m_{0}R$ is \emph{always} larger than the value defined by Eq. (\ref{relation1}).
In summary: we have found that there are two relevant
cases for the chameleon mass. In the first case, the length scale set by
the chameleon mass inside the cavity ($1/m_{0}$) is much smaller than the size of the
experiment. In the second case, the length scale $1/m_{0}$ is ${\cal O}(R)$, and we found an approximate relation Eq.~(\ref{relation1}) between the mass inside the cavity, $R$
and the properties of the potential, encoded here in the power $n$.  Eq. (\ref{relation1}) also defines the smallest possible value of $m_{0}$.  It is clear that for ${\cal O}(1)$ values of $n+1$ we \emph{cannot} have $m_{0} \ll 1/R$.

\section{Chameleon Optics in a Cavity}\label{sec:prop}
It was shown in Ref. \cite{chamPVLAS} that to avoid constraints on solar ALP production
from the CAST experiment, one must require that in bodies with
densities of the order of $10\,\mathrm{g}\cm^{-3}$ the chameleon
mass satisfies $m_{\phi} \geq 10^{4}\,\mathrm{eV}$.  In
experiments such as BRFT, PVLAS,  and Q$\&$A the photon beam typically has
a frequency $\omega = {\cal O}(1)\,\mathrm{eV}$.  In the mirrors then $m_{\phi} \gg \omega$ and so
the chameleon field cannot escape the cavity. Indeed the propagation of the chameleon
field outside the cavity is exponentially attenuated implying that the cavity
mirrors also act as perfect mirrors for the chameleon field. This is at odds with the
usual assumption that ALPs escape from the cavity
without any reflection. The standard expressions for the rotation
(dichroism)  and the ellipticity (birefringence) must therefore be modified.

In this section we study the propagation of a beam of light in the
presence of a chameleon field and derive expressions for the
predicted rotation and ellipticity. For simplicity we initially
assume that there is no distance between the end of the
interaction region and the mirrors, and that the chameleon and
photon fields reflect in the same way.  Whilst neither of these
assumptions are generally true, and there are important effects
associated with the violation of each of them, the calculation is
much simpler and easy to follow if we make them. We say more about
what occurs when these assumptions are dropped in subsection
\ref{sec:prop:phase} below.

We assume that the scalar field mixes with the orthogonal
polarization to the magnetic field. This system is a two-level
system with two states $\vert P> $ and $\vert S>$ for the photon
and the scalar in the absence of the magnetic field. For the
system of differential equations satisfied by the photon and the
scalar, see the Appendices. When a magnetic field is turned on,
the two states mix and the eigenstates are
\begin{eqnarray}
\vert +> &=& \cos \theta \vert S > + i\sin \theta \vert P>\nonumber \\
\vert - > &=& \cos \theta \vert P> + i\sin \theta \vert S > \nonumber
\end{eqnarray}
where
\begin{equation}
\tan 2\theta= \frac{2B\omega}{M m^2}
\end{equation}
The eigenvalues for the above system are given by
\begin{equation}
k_\pm ^2 = \omega^2 -m^2 \frac{\cos 2\theta \pm 1}{2 \cos 2
\theta}.
\end{equation}
For small $\theta$ we get
\begin{equation}
k_+= \omega^2 - m^2 \left( 1+ \frac{\theta^2}{2}\right)
\end{equation}
and
\begin{equation}
k_-^2= \omega^2 + m^2 \theta^2
\end{equation}
In particular we find
\begin{equation}
k_+=\omega - \frac{m^2 }{2\omega}
\label{k1}
\end{equation}
and
\begin{equation}
k_-= \omega+ \frac{m^2\theta^2}{2\omega},
\label{k2}
\end{equation}
which is crucial in the following.

\subsection{Free propagation}
We consider first the situation where the electromagnetic wave and the chameleon propagate freely.  Assuming that the state at a given origin is
$\vert P> (z=0)$, then state at a further position $z$ is given by
\begin{equation}
\vert P>(z)= \cos k_- z \cos \theta \vert -> - i\cos k_+ z
\sin\theta \vert +>
\end{equation}
This mixed state can be expressed in terms of the free scalar and photon as
$$
\vert P>(z)= (\cos^2 \theta \cos k_- z + \sin^2 \theta \cos k_+ z
) \vert P> + i \sin \theta \cos \theta ( \cos k_- z - \cos k_+ z)
\vert S>
$$
The photon part
for small $\theta$ and using the expansion of $k_\pm$ (\ref{k1},\ref{k2}), is
given by
\begin{equation}
\psi(z)=\left(1- 2 \theta^2 \sin^2 \frac{m^2 z}{4\omega}\right)\cos \left( \omega
z + \frac{m^2\theta^2}{2\omega}z-\theta^2 \sin \frac{m^2 z}{
2\omega}\right),
\end{equation}
from which we identity the attenuation and the phase shift
\begin{equation}
a=2 \theta^2 \sin^2 \frac{m^2 z}{4\omega}, \qquad
\delta=\frac{m^2\theta^2}{2\omega}z-\theta^2 \sin \frac{m^2 z}{
2\omega}
\end{equation}
>From those, dividing by two, one gets the rotation and the
ellipticity for an incoming laser beam with a 45 degree
polarization.
We finish the subsection  defining a quantity which will be very
useful below
\begin{equation}
z_{\rm coh}= \frac{2\omega}{m^2},
\end{equation}
measuring the coherence of the system.

\subsection{Propagation in a cavity}

The only difference between the free propagation case and a propagation within a cavity of size $L$ is that we need to sum over
periodic copies shifted by $nL$ .
The mirrors are not perfect, so only $N < \infty$ passes actually occur. Mathematically, this system is equivalent to the coupled wave equations
in a periodic box.

Let us define first
\begin{equation}
a_n(z)=2 \theta^2 \sin^2 \frac{m^2 (z+nL)}{4\omega}
\end{equation}
and
\begin{equation}
\delta_n(z)=\frac{m^2\theta^2}{2\omega}(z+nL)-\theta^2 \sin
\frac{m^2 (z+nL)}{ 2\omega}
\end{equation}
The photon wave function is then given by summing
\begin{equation}
\psi (z)=\sum_{n=0}^{N-1} (1- a_n)\cos k_- (z+nL) + \theta^2 \sin
\frac{m^2 (z+nL)}{ 2\omega} \sin k_- (z+nL)
\end{equation}
 In a perfect
cavity, the waves have a resonance
, i.e. $\omega L=2\pi p$. Therefore
\begin{eqnarray}
\psi (z)=&&\sum_{n=0}^{N-1} (1- a_n)\cos \left(\omega  z + \frac{m^2
\theta^2}{2\omega}(z+nL)\right) \\ &&+ \theta^2 \sin \frac{m^2 (z+nL)}{
2\omega} \sin k_- (z+nL) \nonumber
\end{eqnarray}
Using the small $\theta$  approximation, one obtains
\begin{eqnarray}
\psi (z)=&&N \left(1- \frac{1}{N} \sum_{n=0}^{N-1} a_n\right)\cos \omega z
\\ &&+\sum_{n=0}^{N-1} \left(\theta^2 \sin \frac{m^2 (z+nL)}{ 2\omega}-
\frac{m^2\theta^2}{2\omega}(z+nL)\right) \sin \omega z \nonumber
\end{eqnarray}
Notice this is nothing but a standing wave
\begin{equation}
\psi(z)=N\left(1- \frac{1}{N} \sum_{n=0}^{N-1} a_n(z)\right)\cos \left( \omega z +
\frac{1}{N} \sum_{n=0}^{N-1} \delta_n(z)\right)
\end{equation}
So the phase shift and the attenuation at the end of the cavity $z=L$ are given by
\begin{equation}
\delta_T=\frac{1}{N} \sum_{n=0}^{N-1} \delta_n(L), \qquad a_T=\frac{1}{N} \sum_{n=0}^{N-1} a_n(L)
\end{equation}

These sums can be exactly computed, which give
\begin{equation}
\delta_T = \theta^2\left (\frac{(N+1)}{2} \frac{L}{z_{\rm coh}} +
\frac{ \sin N \frac{L}{z_{\rm coh}} + \sin \frac{L}{z_{\rm coh}}
-\sin (N+1) \frac{L}{z_{\rm coh}}}{2N(1-\cos
\frac{L}{z_{coh}})}\right )
\end{equation}
For
larger values and using  $\sin x \le x$, we have the inequality
$\delta_T \le \delta_{N}$.

Similarly the attenuation is
\begin{equation}
a_T= \theta^2 \left (1 - \frac{ 1+ \cos N \frac{L}{z_{\rm coh}} -
\cos \frac{L}{z_{\rm coh}} - \cos (N+1) \frac{L}{z_{\rm
coh}}}{2N(1- \cos \frac{L}{z_{\rm coh}})}\right).
\end{equation}


\subsection{Coherence}

The previous expressions can be easily interpreted in terms of the
coherence of the photon-scalar system.  We focus on the case where
the coherence length and the length of the experiment are
commensurate. Define the number of coherent passes as $P L= 2\pi
z_{\rm coh}$. Using this one finds that
\begin{equation}
a_{n+P}= a_n, \qquad \delta_{n+P}= \delta_n + \frac{m^2\theta^2}{2\omega} PL
\end{equation}
The attenuation is then given by
\begin{equation}
a_T= \frac{1}{N} \sum_{j=0}^{N/P-1} \sum_{n=0}^{P-1} a_{n+
jP}=\frac{1}{N} \sum_{j=0}^{N/P-1} \sum_{n=0}^{P-1} a_{n}
\end{equation}
and therefore
\begin{equation}
a_T=\frac{1}{P}\sum_{n=0}^{P-1} a_{n}
\end{equation}
Hence the attenuation depends only on the coherence length, after
$P$ passes, the waves are not coherent any more. Using the previous
formulae we find
\begin{equation}
a_T=\theta^2
\end{equation}
and for the rotation
\begin{equation}
{\rm rotation /\rm pass}= \frac {\theta^2}{2N}
\end{equation}
which was first derived in \cite{zav}.

In order to get the ellipticity one notes first that the phase shift is given by
\begin{equation}
\delta_T=\frac{1}{N} \sum_{j=0}^{N/P-1} \sum
_{n=0}^{P-1}\delta_{n+jP}=\frac{1}{N} \sum_{j=0}^{N/P-1} \sum
_{n=0}^{P-1}\left(\delta_{n}+ jPL \frac{m^2\theta^2}{2\omega}\right)
\end{equation}
and therefore
\begin{equation}
\delta_T= PL (\frac{N}{P}-1) \frac{m^2 \theta^2}{4\omega}+
(P+1)\theta^2 \frac{L}{2 z_{\rm coh}}
\end{equation}
Using the definition of $P$ one finds
\begin{equation}
\delta_T= \pi  \left(\frac{N}{P}-1\right) \theta^2+ 2\pi
\frac{P+1}{P}\theta^2
\end{equation}
For large $P$  this simplifies to
\begin{equation}
\delta_T= \pi\left(\frac{ N}{P}-1\right) \theta^2
\end{equation}
This can also be written for $N/P \gg 1$ as
\begin{equation}
\delta_T= \frac{NL m^2}{4\omega} \theta^2
\end{equation}
Which again was found in \cite{zav}. The ellipticity per pass is now
\begin{equation}
{\rm ellipticity/ pass}= \frac{ \pi \theta^2}{2}
\left(\frac{1}{P}-\frac{1}{N}\right)
\end{equation}
and one finds that the  ellipticity per rotation is
\begin{equation}
\frac{\rm ellipticity}{\rm rotation}= \pi \left(\frac{
N}{P}-1\right)= \pi\left(\frac{NL}{2\pi z_{\rm coh}}-1\right)
\label{ratio}
\end{equation}
 In particular, for large N, the ellipticity is always much
larger than the rotation. This fact is still true when the
non-interacting zone of length $d$ is taken into account. It is a
crucial prediction of chameleon theories.

\subsection{Phase Shifts}
\label{sec:prop:phase} The above calculation has been performed
under the assumptions that there is no gap between the ends of the
interaction region and the mirrors and that the chameleon field
reflects off the mirror in precisely the same way as the photon
does.  Unfortunately, neither of these assumptions are generally
true.

In the PVLAS, BRFT and Q$\&$A experiments, there is always some gap
between the end of the interaction region and the mirrors. In
PVLAS this gap is $270\cm$ long.  Outside the interaction region both the chameleon field
and the photon field propagate freely, however since the chameleon
field is massive it travels more slowly than the photon.  As we
show in  Appendix \ref{appB}, this leads to the chameleon field picking up
a phase shift $\Delta_{m}$ relative to the photon field when the field returns to the interaction region. This phase shift alters the
formulae for the rotation and ellipticity. We find that if there
is a distance $d$ between the mirror outside the interaction
region, then $\Delta_{m} \approx m^2_{\phi} d/\omega$.
\begin{figure}[tbh]
\begin{center}
 \includegraphics[width=8.5cm]{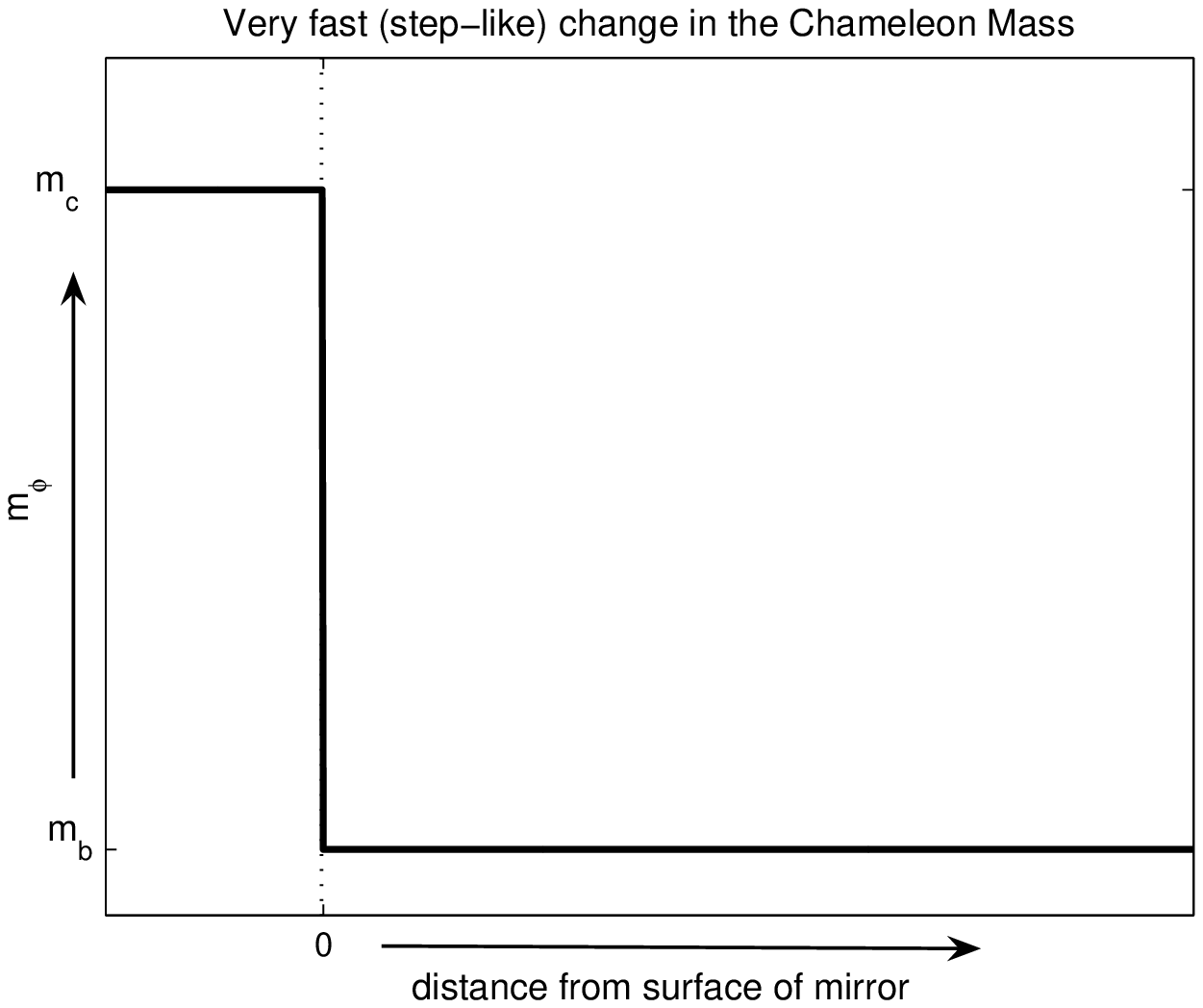}
\includegraphics[width=8.5cm] {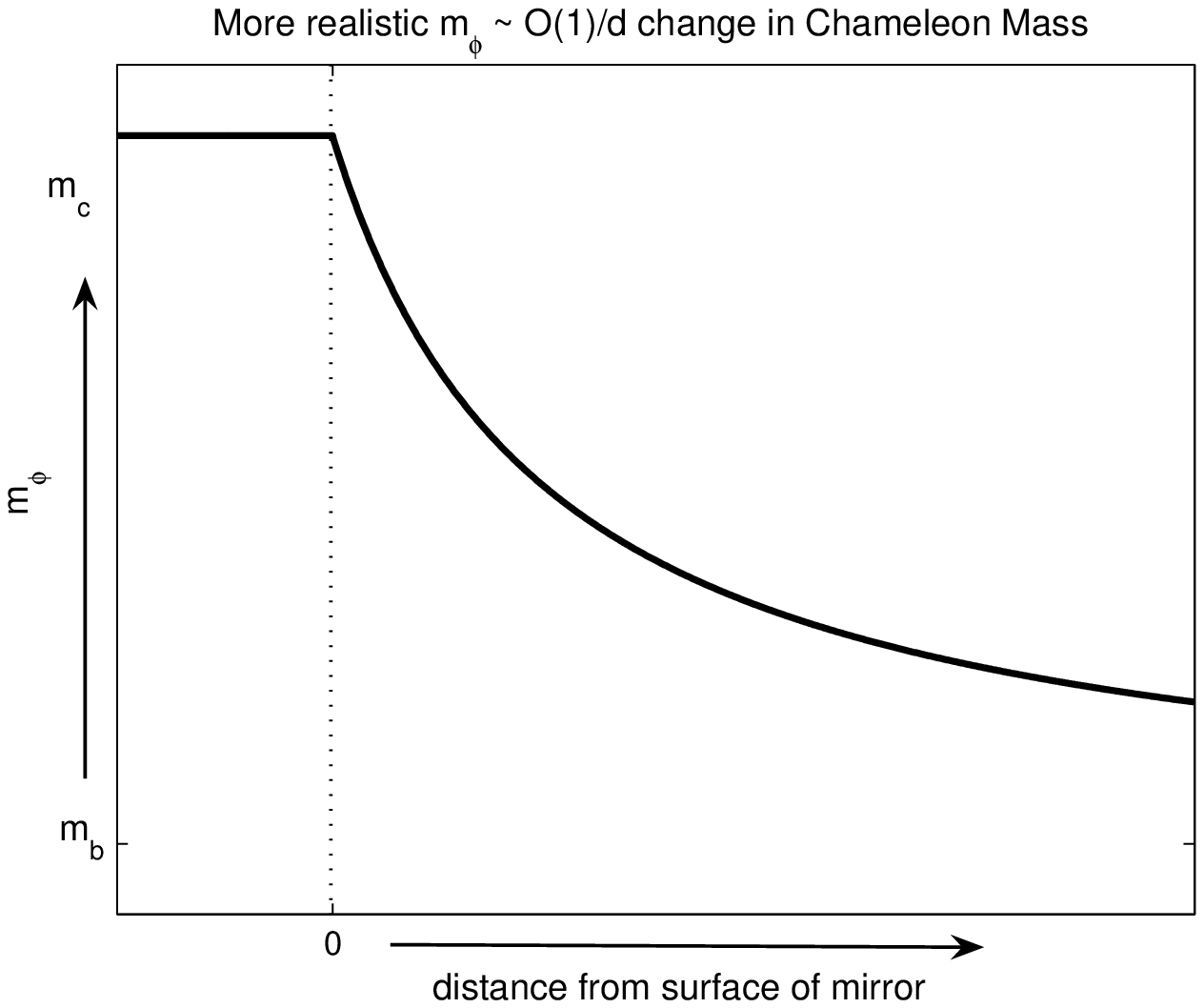}
\end{center}
\caption{Illustration of the difference between a sharp (step-like) change in the chameleon
mass at the surface of the mirror and the more realistic $m_{\phi} \sim {\cal O}(1/d)$,
for $d \gg 1/m_{c}$, behaviour, where $d$ is the distance from the surface of the mirror.
In both of these sketches, $m_{c}$ is the chameleon mass inside the mirror and $m_{b}$
is the chameleon mass far from the mirror. The dotted line indicates the surface of the mirror.}
\label{fig1}
\end{figure}
In addition to the phase shift $\Delta_{m}$, there is generally an
additional phase shift $\Delta_{r}$ due to the manner in which the
chameleon field reflects off the mirror.  Consider a mirror placed
at $x =0$.  If $m_{\phi}$ behaved liked a step function, see Figure
\ref{fig1}, where $m_{\phi} = m_{<}$ for $x < 0$ and $m_{\phi} =
m_{>}$ for $x > 0$ and $m_{>} \ll m_{<}$, then the wave function
of a chameleon wave with wave-number $k$ incident on the mirror would, in $x>0$,
have the form:
$$
\delta \phi \approx \delta\phi_{0}\sin(k x).
$$
When $x < 0$, $\phi$ decays away exponentially.    However, in
realistic theories $\phi$ is continuous across $x=0$ and so
$m_{\phi}$ is too.  Indeed, in Appendix \ref{appB}, we show, via a similar
calculation to the one performed in Section \ref{sec:cylinder}
above, that for $1/m_{<} \lesssim x \lesssim 1/m_{>}$ we have
$m_{\phi} \approx {\cal O}(1)/x$ (see Figure \ref{fig1}).

It should be noted that this behaviour is not specific to the
inverse power-law potentials considered here, but is generic to
almost all chameleon theories (see  Appendix \ref{appB} for a complete
discussion).    A scalar wave with frequency $\omega$ switches
from oscillatory to exponentially decaying behaviour when
$m_{\phi}^2 = \omega^2$.  In chameleon theories this transition
therefore occurs at a distance $x_{r} = {\cal O}(1/\omega)$ from the
mirror. Chameleon waves therefore behave as if they reflect not at
$x=0$ but at $x=x_{r} > 0$.  Schematically this alters the form of
the chameleon wave function in $x > x_{r}$ to:
$$
\delta \phi \approx \delta\phi_{0}\sin(k (x-x_r)).
$$
As is shown in Appendix \ref{appB}, this early reflection produces another phase shift, $\Delta_{r}$, in the chameleon wave relative to the photon wave when they return to the interaction region.  We find that for inverse square law potentials:
$$
\Delta_{r} = \frac{\pi n}{n+2},
$$
where $n > 0$.

The presence of a total phase shift $\Delta \approx \Delta_{r} +
\Delta_{m}$ alters the formulae for the rotation, $\Delta \varphi$
and ellipticity, $\psi$ and these are calculated in Appendix
\ref{appA}. Both of the mentioned effects  will exist, to some
degree in all theories in which the ALP does not escape the
cavity.  We find that when $\Delta \neq 0$ and provided that:
\begin{equation}
\left(\frac{B\omega \sin\left(m_{\phi}^2 L /4\omega\right)}{M m_{\phi}^2 \sin\left(\Delta/2 + m_{\phi}^2 L /4\omega\right)}\right)^2 \ll 1,
\end{equation}
we predict
\begin{eqnarray}
\frac{\Delta \varphi}{\sin 2 \varphi} =&&-\left(\frac{B\omega}{M m_{\phi}^2}\right)^2 H_{\Delta}\left(\frac{m_{\phi}^2 L}{2\omega}\right)\left\lbrace \frac{1}{2} + \left[\sin^2 \left(\frac{N \Delta}{2} + \frac{Nm_{\phi}^2 L}{4\omega}\right)\right.\right. \label{rotpredicteqn} \\&&-\left.\left.\frac{1}{2}\right]\delta_{N}\left(\Delta+ \frac{m_{\phi}^2L}{2\omega}\right)\right\rbrace, \nonumber
\end{eqnarray}
and
\begin{eqnarray}
\frac{\psi}{\sin 2 \varphi} = && -\frac{1}{2}\left(\frac{B \omega}{M m_{\phi}^2}\right)^2\left\lbrace\frac{Nm^2_{\phi} L}{2\omega} - NG_{\Delta}\left(\frac{m_{\phi}^2 L}{2\omega}\right) \right. \label{ellippredicteqn} \\ &&-\left. \sin\left(N\Delta + \frac{Nm_{\phi}^2 L}{2\omega}\right)H_{\Delta}\left(\frac{m_{\phi}^2 L}{2\omega}\right)\delta_{N}\left(\Delta+ \frac{m_{\phi}^2L}{2\omega}\right)\right\rbrace. \nonumber
\end{eqnarray}
where
\begin{eqnarray}
G_{\Delta}(x) &=& \frac{2\sin(\Delta/2)\sin(x/2)}{\sin(\Delta/2 + x/2)}, \\
H_{\Delta}(x) &=& \frac{\sin^2(x/2)}{\sin^2(\Delta/2 + x/2)},
\end{eqnarray}
and
\begin{equation}
\delta_{N}(x) = \frac{\sin((N+1)x)}{(N+1)\sin x}.
\end{equation}
The phase $\Delta$ is given by
\begin{equation}
\Delta = \frac{m_\phi^2 d}{\omega} + \Delta_r,
\end{equation}
where $\Delta_r$ depends on the potential $V(\phi)$. For inverse power-law potential ($n>0$), $\Delta_r$ is given by
\begin{equation}
\Delta_r = \frac{\pi n}{n+2}.
\end{equation}
When $\Delta = 0$, $G_{\Delta} = 0$ and $H_{\Delta} = 1$.  We always
have $\vert \delta_{N}(x) \vert < 1$, and for $N$ large
$\delta_{N}(x)$ is strongly peaked about $x=m \pi$ for any integer
$m$ and $\delta_{N}(x) \ll 1$ otherwise. In many situations one
finds that $m_{\phi}^2 (L/2 + d)/2\omega \ll 1$ and
$\tan(\Delta_{r}/2) \sim {\cal O}(1)$.  In these circumstances, the
expressions for the ellipticity and rotation simply greatly:
\begin{eqnarray*}
\frac{\Delta \varphi}{\sin \varphi} &\approx& \frac{B^2 L^2}{32 M^2 \sin^2\left(\frac{\Delta_{r}}{2}\right)},\\
\frac{\psi}{\sin \varphi} &\approx& -\frac{N B^2 L^2}{16 M^2 \tan\left(\frac{\Delta_{r}}{2}\right)},
\end{eqnarray*}
which are both independent of $m_{\phi}$. In this limit, the ratio
of the rotation to the ellipticity is:
\begin{eqnarray}
\frac{\Delta \varphi}{\psi} = \frac{1}{N\sin(\Delta_{r})}. \label{ratiofull}
\end{eqnarray}
\section{Predictions}\label{sec:predictions}
Having described the physics and derived the basic formulae for
rotation and ellipticity induced by the coupling of a chameleon
field to the electromagnetic sector, we now study the predictions of
the chameleon model. To simplify the analysis, we have assumed that
the chameleon couples to all matter types (including photons) with
the same strength. In this case, there is a lower bound on the
energy scale $M$ coming from the contribution of the chameleon to
the anomalous magnetic moment of the muon and electron. The
contribution is of order $(m_{e,\mu}/M)^2$ \cite{carlson}. In order
for the chameleon contribution to the anomalous magnetic moment of
the muon to be small enough, $M$ has to be bigger than $M\approx
10^4$~GeV. We will therefore concentrate here on the case
$M>10^4$~GeV. We should mention here that if the couplings differ
from species to species, this constraint on the coupling to the
electromagnetic sector could be relaxed.

\begin{figure}[htb!]
\includegraphics*[width=12cm]{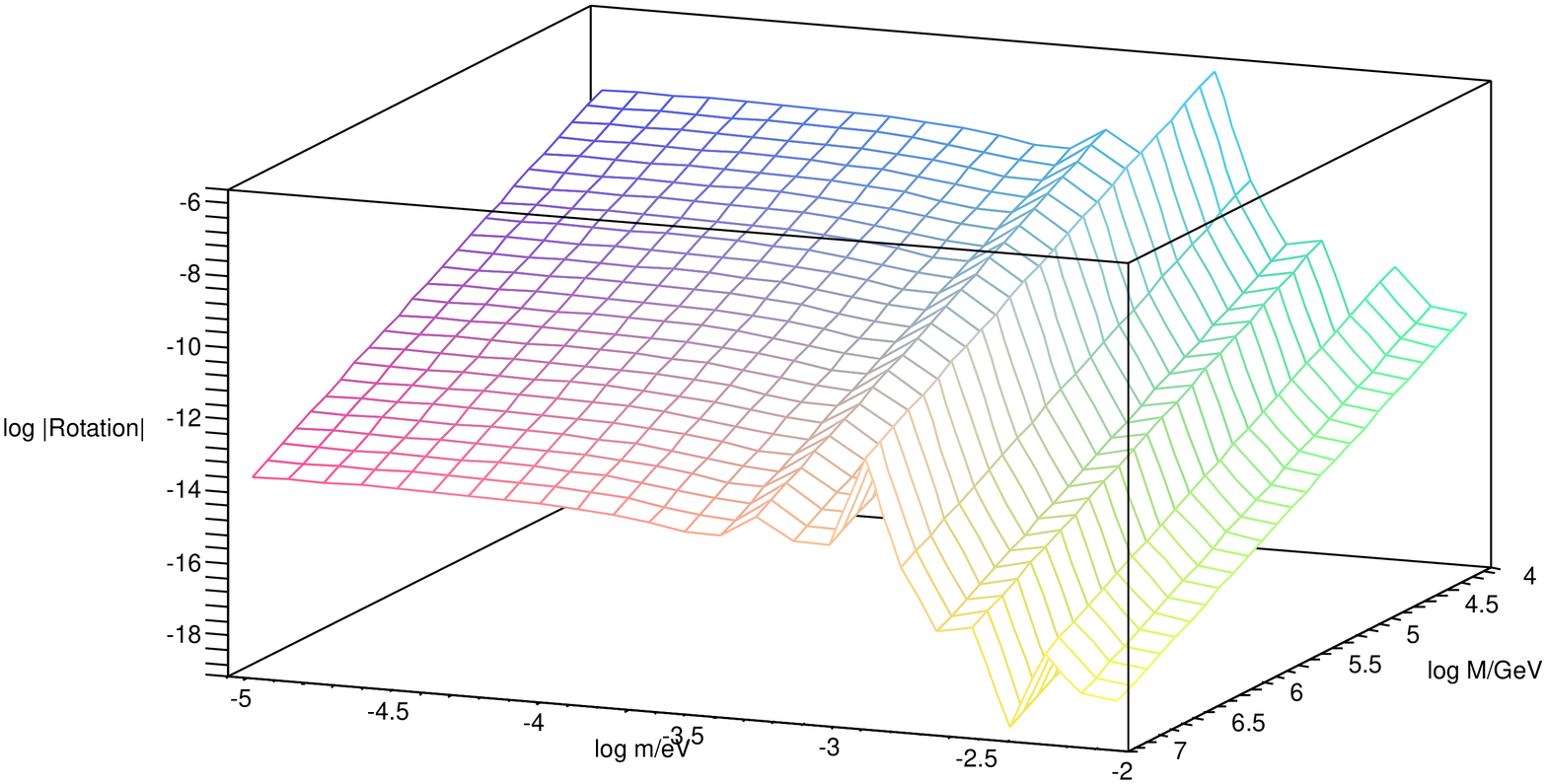}
\includegraphics*[width=12cm]{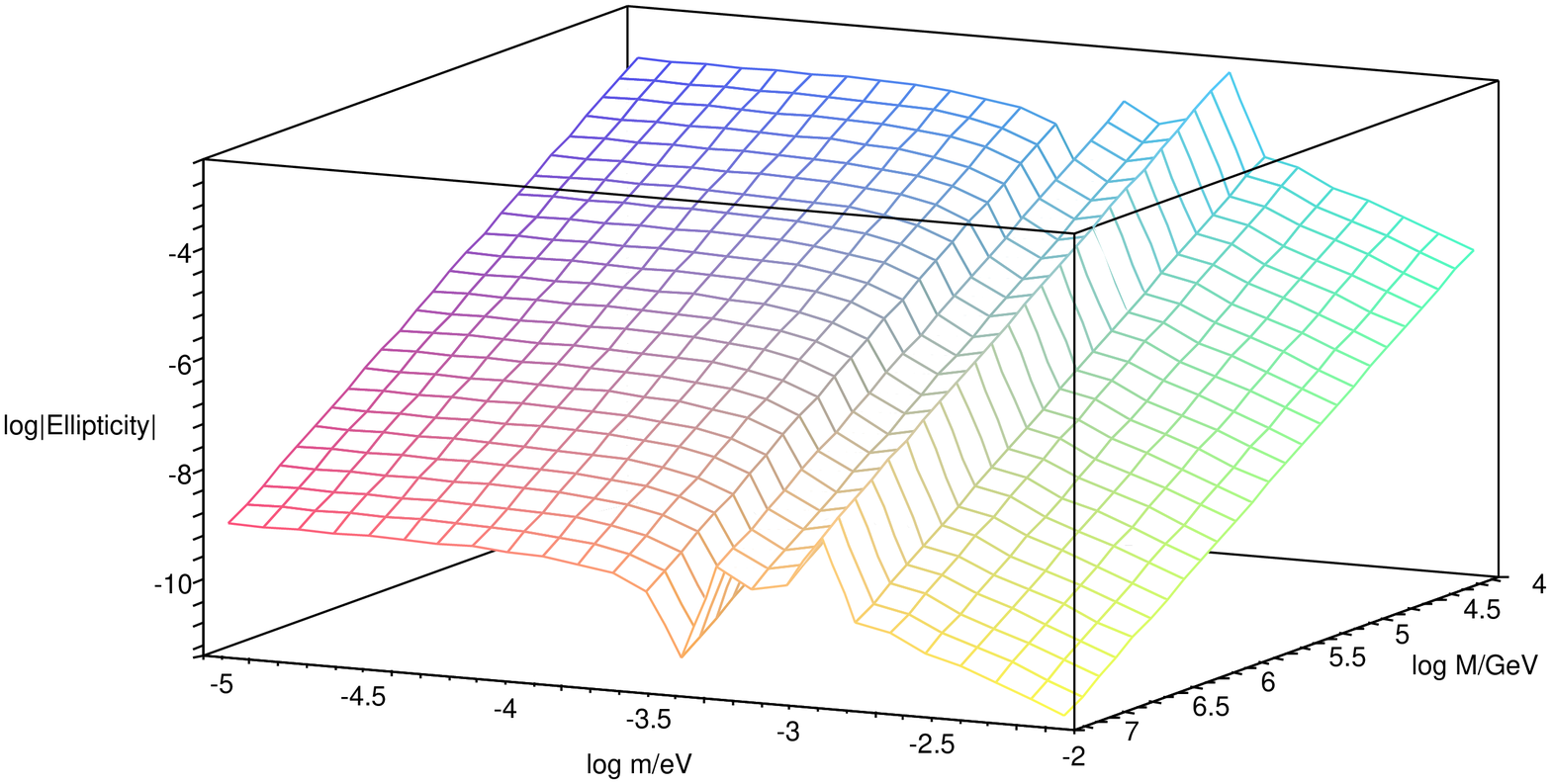}
\caption[]{Predictions for rotation (upper plot) and ellipticity (lower plot) in the chameleon model as a function of $M$ and $m_\phi$. The PVLAS set-up is used
($L=100$~cm, $d=270$~cm, $\omega=1.2$eV, $B=5$~T and $\varphi=\pi/4$). Furthermore we have chosen $n=1$.}
\label{fig:rotellip}
\end{figure}

The predictions for rotation and ellipticity are shown in Figure \ref{fig:rotellip}. One feature of the chameleon model is that the ellipticity is generally predicted to be much larger than the rotation. This can be viewed as a
generic prediction of chameleon theories. In general, a large
ellipticity to rotation ratio occurs (for large $N$) whenever:
\begin{itemize}
\item The ALPs are reflected by the mirrors rather escaping through them.
\item The ALPs from previous passes are not coherent with the photon field.
\end{itemize}
If the ALPs were to escape then each pass would give the same
contribution to the rotation and ellipticity, these add up and
thus both $\Delta \varphi$ and $\psi$ would be proportional to
$N$.  However, if the first of the above conditions is satisfied
then this is not the case, since ALPs from previous passes
interact with the photon field and thus the way in which the
rotation and ellipticity build up is altered. If, additionally,
ALPs from previous passes are \emph{not} coherent with the photon
field, then a large degree of cancellation of both the ellipticity
and rotation occurs.  For large $N$, the cancellation in the
rotation, $\Delta \varphi$, is almost exact, and as result $\Delta
\varphi$ is almost independent of $N$.

Cancellation of contributions to the ellipticity also occurs. However,
there is
always at least one contribution which does not cancel but builds up
as the number of passes increases.  This contribution results from
the fact that the component of the photon field that interacts
with the ALP propagates more slowly in the interaction region than
its non-interacting component. As a result the relative phase of
the interacting and non-interacting components is shifted. This
contribution to the ellipticity is unaffected by the decoherent
ALPs.  There is, therefore, always at least one contribution to
the ellipticity that is proportional to $N$.  Thus the combination
of reflecting ALPs and their becoming decoherent results in the
ellipticity but not the rotation growing as $N$. The rotation to
ellipticity ratio is therefore $\sim 1/N \ll 1$ (see Eq. (\ref{ratiofull})
for a better estimate).

If the ALP in question is a chameleon field then both
reflection and decoherence almost certainly occur.  Reflection
occurs because the mass of the chameleon inside the mirror, $m_c$,
is much larger than the mass far outside, $m_b$, and the energy of
beam, $\omega$, used in experiments is almost always $\ll m_{c}$.
Decoherence occurs because, as we found in  Appendix \ref{appB}, at
a distance $x$ ($1/m_c \ll x \ll 1/m_b$) from the surface of the
mirror, the chameleon mass behaves as $\sim {\mathcal{O}}(1)/x$.
The reflection of the chameleon does not, therefore, occur on the
surface of the mirror but a distance $x_{r} \sim
{\mathcal{O}}(1)/\omega$ from the mirror's surface. This, we
found, resulted in an ${\mathcal{O}}(1)$ phase shift between the
reflected chameleon and photon fields. Generally then, the
chameleon fields from previous passes are both reflected and are
not coherent with the photon field.  Whilst, one could probably
construct a chameleon theory where one or both of the above
conditions did not hold, it would not be particularly generic and
might also have problems satisfying gravitational and other
laboratory constraints on such theories.

It is therefore a generic feature of chameleon theories that the
rotation to ellipticity ratio is $O(1/N) \ll 1$.  This means that,
according to the chameleon model, it is easier to detect the
ellipticity than the rotation.

In Section \ref{sec:cylinder} we showed how, for a given experimental set-up
and choice of $V(\phi)$, the chameleon mass, $m_{\phi}$, in the interaction
could be calculated from the chameleon-to-matter coupling if $\Lambda$
and $n$ are known. We found that there were two regimes.  If the density of
the vacuum, $\rho_{\rm gas}$, and the radius, $R$, of the cavity were small
enough and $M$ large enough then $m_{\phi} \sim {\mathcal{O}}(2)/R$ for
small $n$. Alternatively, if $M$ is small enough or $\rho_{\rm gas}$ / $R$ large
enough, then the chameleon mass depends on $(\rho_{\rm gas}+B^2/2)/M$. The
chameleon mass and therefore the predicted ellipticity and rotation are
therefore highly dependent on the set-up of the axion search experiment.
In the chameleon model,  there is no reason to expect that two different
experiments should detect the same ellipticity and rotation.

\subsection{Predictions for PVLAS, Q$\&$A, BMV and BRFT}
\begin{figure}[htb!]
\includegraphics*[width=8.5cm]{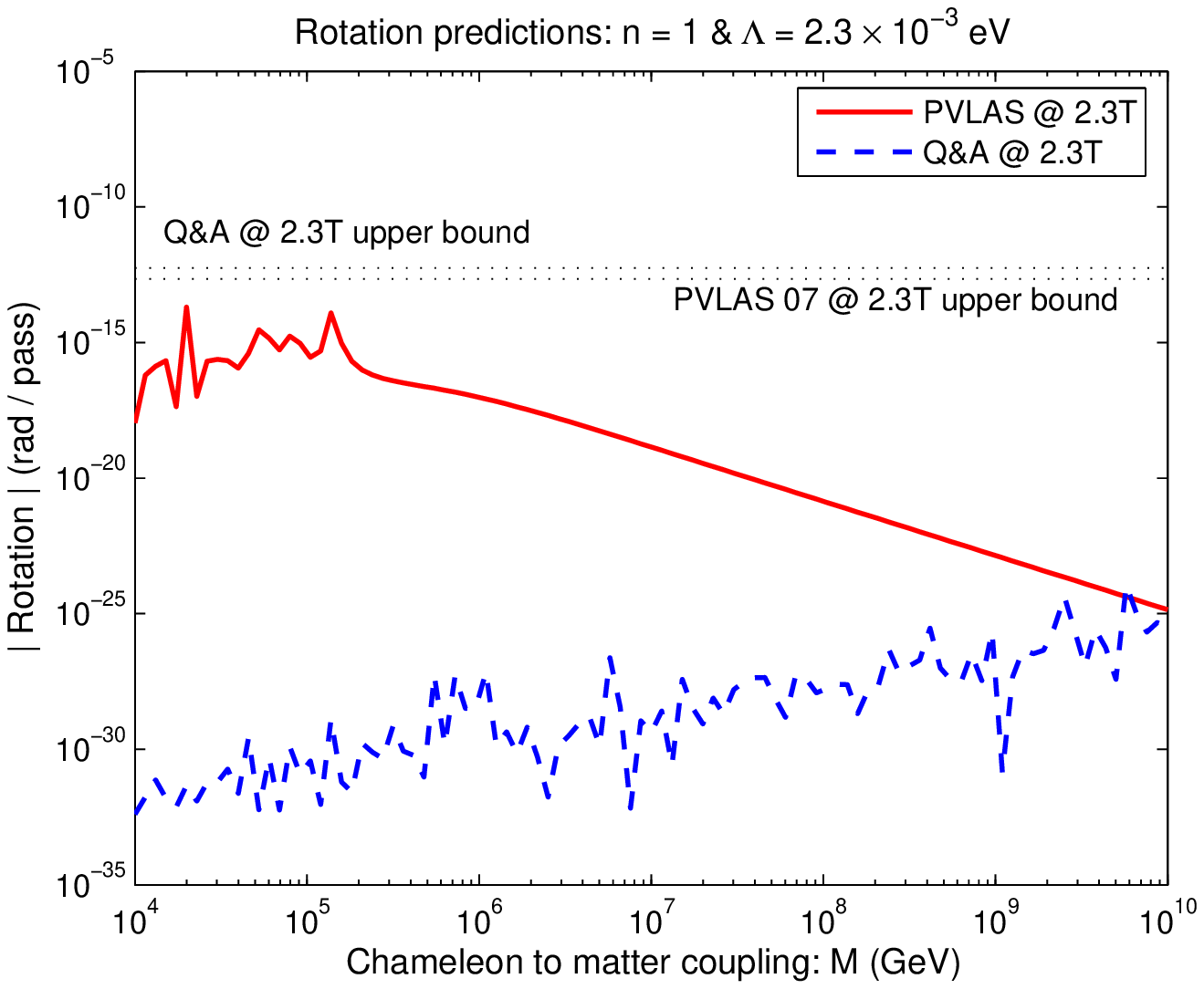}
\includegraphics*[width=8.5cm]{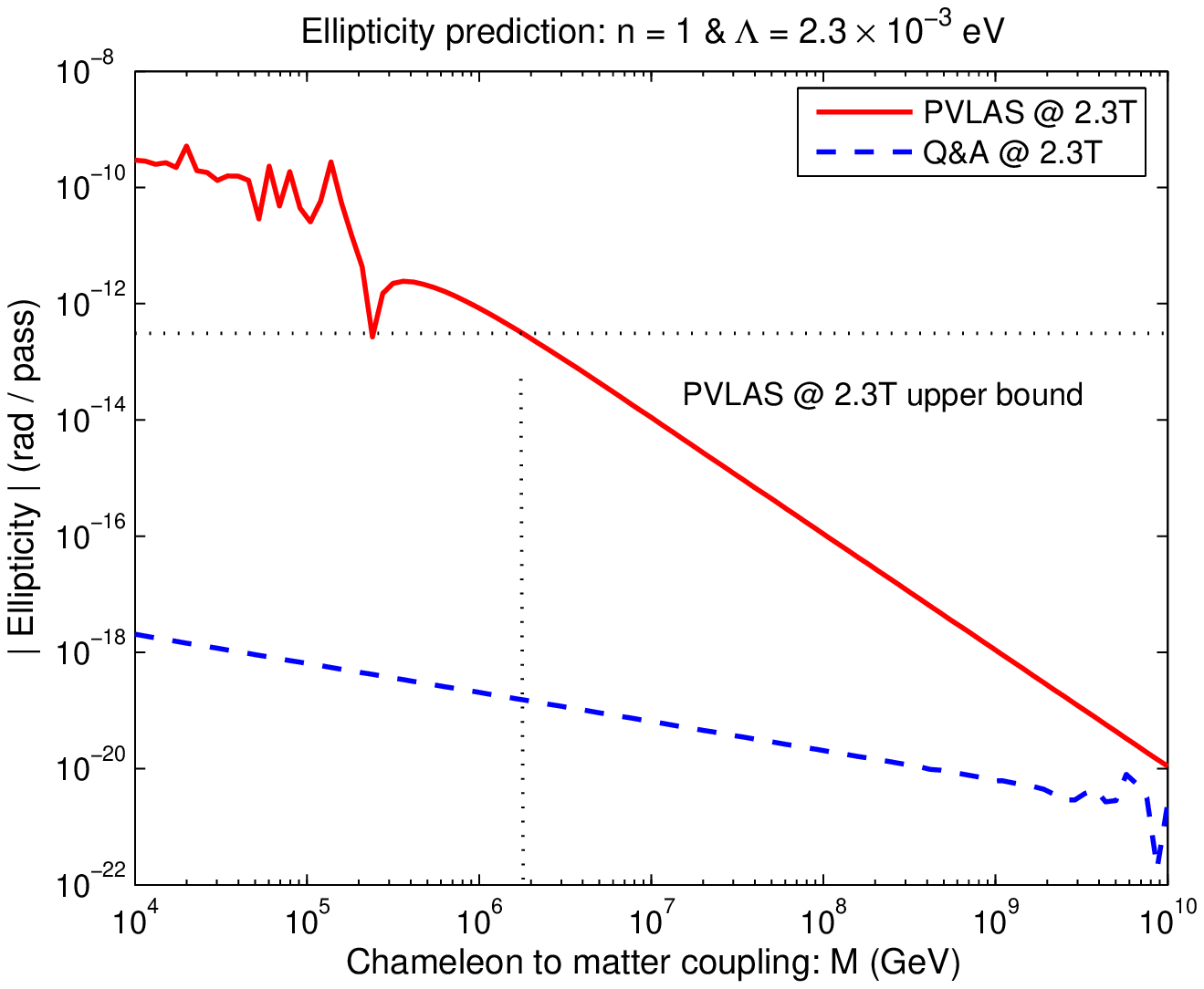}
\caption[]{Predictions for rotation (left) and ellipticity (right) in the chameleon model as a function of $M$ for $\Lambda = 2.3 \times 10^{-3}\,{\rm eV}$ and $n=1$. Predictions for the $2.3$~T PVLAS ($L=100$~cm, $d=270$~cm, $\omega=1.2$eV, $B=2.3$T, $\rho_{\rm gas} = 2\times 10^{-14}{\rm g cm}^{-3}$ and $\varphi=\pi/4$) and Q$\&$A ($L=60$~cm,$d=145$~cm, $\omega=1.2$eV, $B=2.3$T, $\rho_{\rm gas} = 8.5 \times 10^{-9}{\rm g cm}^{-3}$ and $\varphi=\pi/4$) set-ups are shown.  The thin-dotted lines show the $95\%$ confidence upper bounds on both the rotation and the ellipticity. }
\label{fig:PVLASQA}
\end{figure}
As we argued in Section \ref{sec:chammodel}, we expect
$\Lambda \approx 2.3 \times 10^{-3}\,{\rm eV}$ if the energy density of
the chameleon field is to be associated with dark energy.  Ideally, one
would determine both $\Lambda$ and $n$ from an experimental detection.
However, since the status of the only (i.e. PVLAS) detection reported to date
is unclear, we now consider the specific predictions for the chameleon model
for $\Lambda \approx 2.3 \times 10^{-3}\,{\rm eV}$ and $n=1$.  The vacuum used
in the PVLAS set-up is very good, with a density of about
$2 \times 10^{-14}\,{\rm g cm}^{-3}$, whereas the density of the Q$\&$A
experiment's vacuum is significantly higher: $8.5 \times 10^{-9}\,{\rm g cm}^{-3}$;
as a result, for $M \sim 10^{6} - 10^{10} \GeV$, the chameleon mass in the
PVLAS experiment $\sim {\mathcal{O}}(2)/R$ and hence independent of $M$.
In the Q$\&$A experiment, however, $m_{\phi}$ is, for given $M$, both larger
than it is in the PVLAS set-up and depends on $(\rho_{\rm gas}+B^2/2)/M$. The
larger value of $m_{\phi}$ in the Q$\&$A set-up means that the ellipticity and
rotation predicted by the chameleon model are far smaller than those predicted
for the PVLAS experiment. The chameleon model predictions for both the PVLAS
and Q$\&$A set-up are shown in Figure \ref{fig:PVLASQA}.  The thin dotted lines
in these plots show the $2.3$~T PVLAS 2007 \cite{PVLAS07} and Q$\&$A $95\%$
confidence limits on the rotation and ellipticity. For $M > 10^{4}\GeV$ it
is only the $2.3$~T PVLAS upper bound on the ellipticity that provides a
useful constraint on $M$: we must require $M \gtrsim 2 \times 10^{6}\GeV$
if $n=1$ and $\Lambda \approx 2.3 \times 10^{-3}\eV$.  A similar limit
of $M$ is found for other values of $n$.

\begin{figure}[htb!]
\includegraphics*[width=8.5cm]{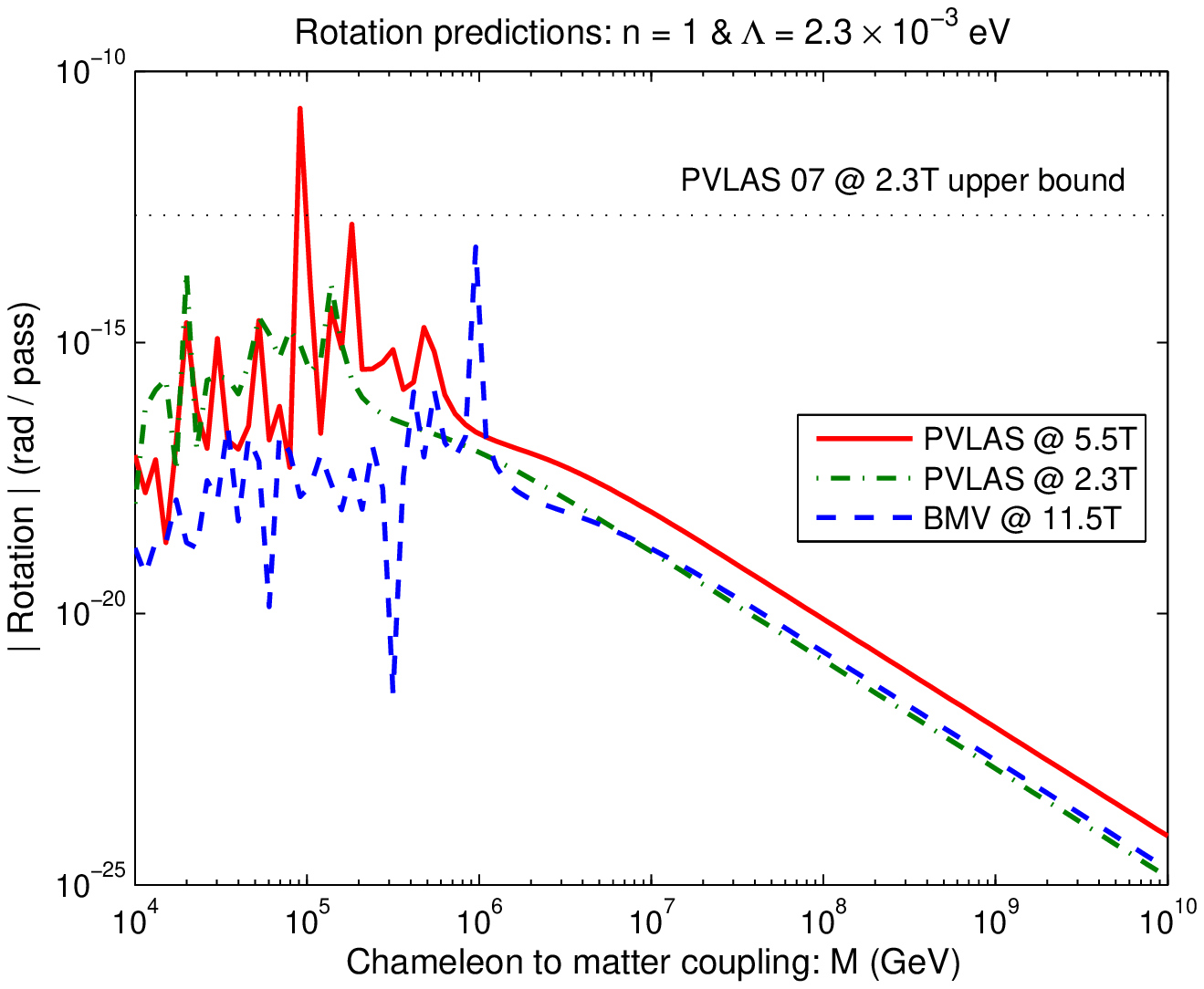}
\includegraphics*[width=8.5cm]{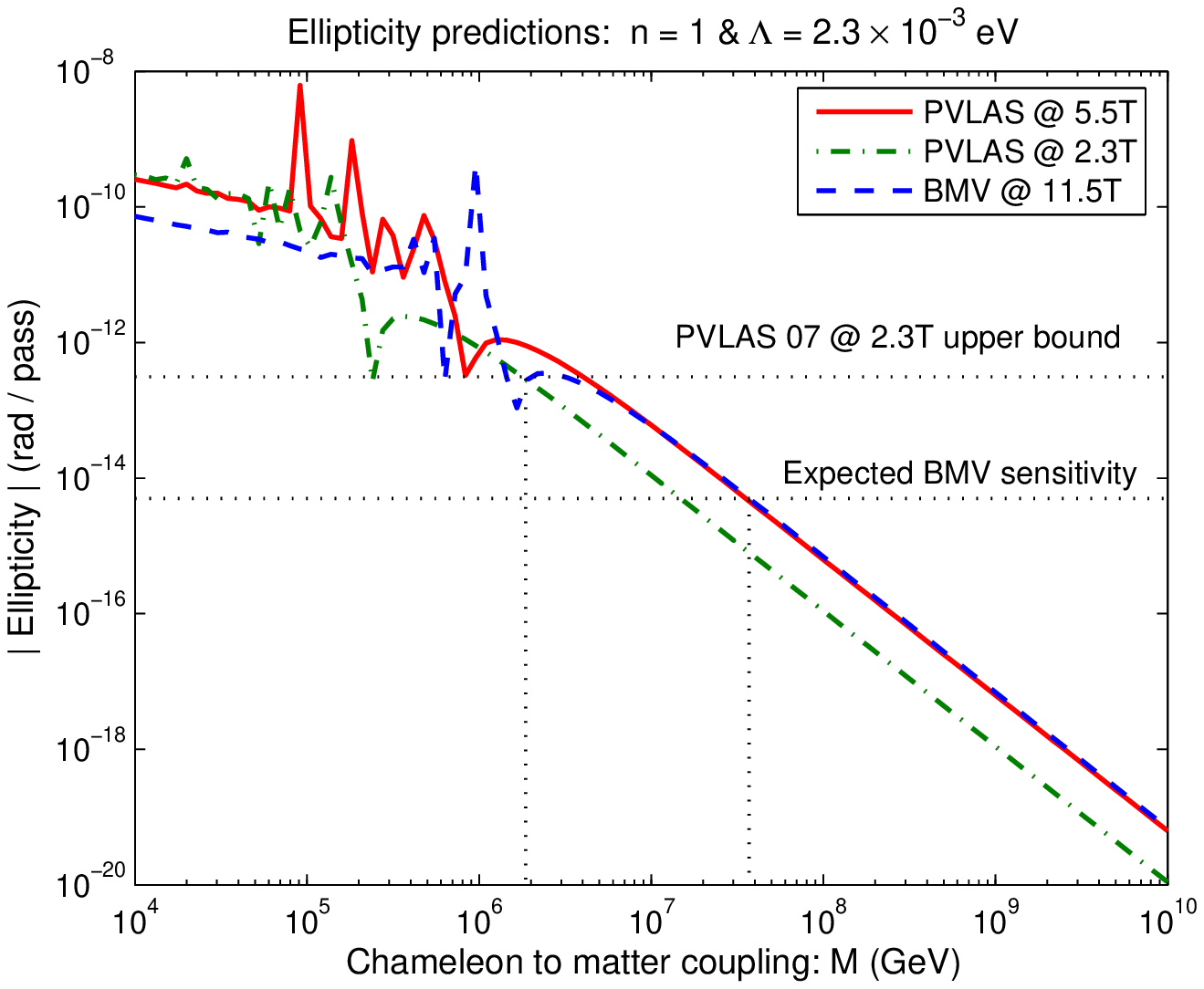}
\caption[]{Predictions for rotation (left) and ellipticity (right) in the chameleon model as a function of $M$ for $\Lambda = 2.3 \times 10^{-3}\,{\rm eV}$ and $n=1$. Predictions for the $B=2.3$~T and $B=5.5$~T PVLAS ($L=100$~cm, $d=270$~cm, $\omega=1.2$~eV, $\rho_{\rm gas} = 2\times 10^{-14}{\rm g cm}^{-3}$ and $\varphi=\pi/4$) and BMV ($L=50$~cm,$d=85$~cm, $\omega=1.2$~eV, $B=11.5$~T, $\rho_{\rm gas} \approx 10^{-14}{\rm g cm}^{-3}$ and $\varphi=\pi/4$) set-ups are shown.  The thin-dotted lines show the $95\%$ confidence upper bounds on both the rotation and the ellipticity. }
\label{fig:PVLASBMV}
\end{figure}
The PVLAS experiment performs better than Q$\&$A as a probe for
chameleon fields because of the high quality vacuum it employs.
The upcoming BMV experiment uses a similar high quality vacuum
with pressure $< 10^{-8}\,{\rm mbar}$ \cite{BMV,BMV1}. This experiment
will additionally use a higher strength magnetic field than PVLAS,
employ a greater number of passes and additionally promises a
higher precision.  Figure \ref{fig:PVLASBMV} shows the predicted
rotation and ellipticity signals for PVLAS (with both $B=2.3$~T
and $5.5$~T) and BMV again with $n=1$ and $\Lambda = 2.3 \times
10^{-3}\eV$.  We see that BMV should be able to detect, or rule
out, such chameleon theories with $M \lesssim 3 \times 10^{7}
\GeV$ which represents an order of magnitude improvement over
PVLAS.

\begin{figure}[htb!]
\includegraphics*[width=8.5cm]{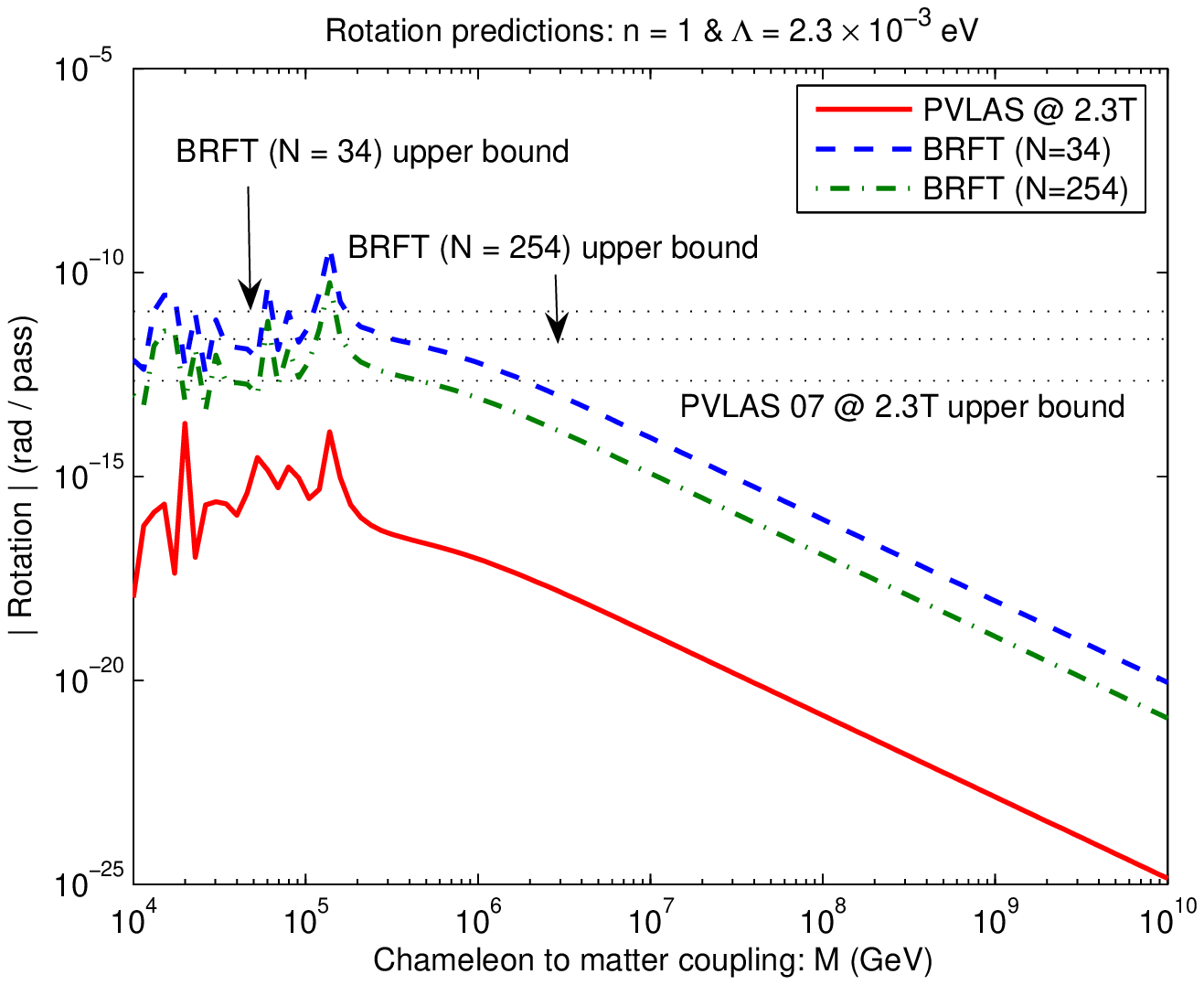}
\includegraphics*[width=8.5cm]{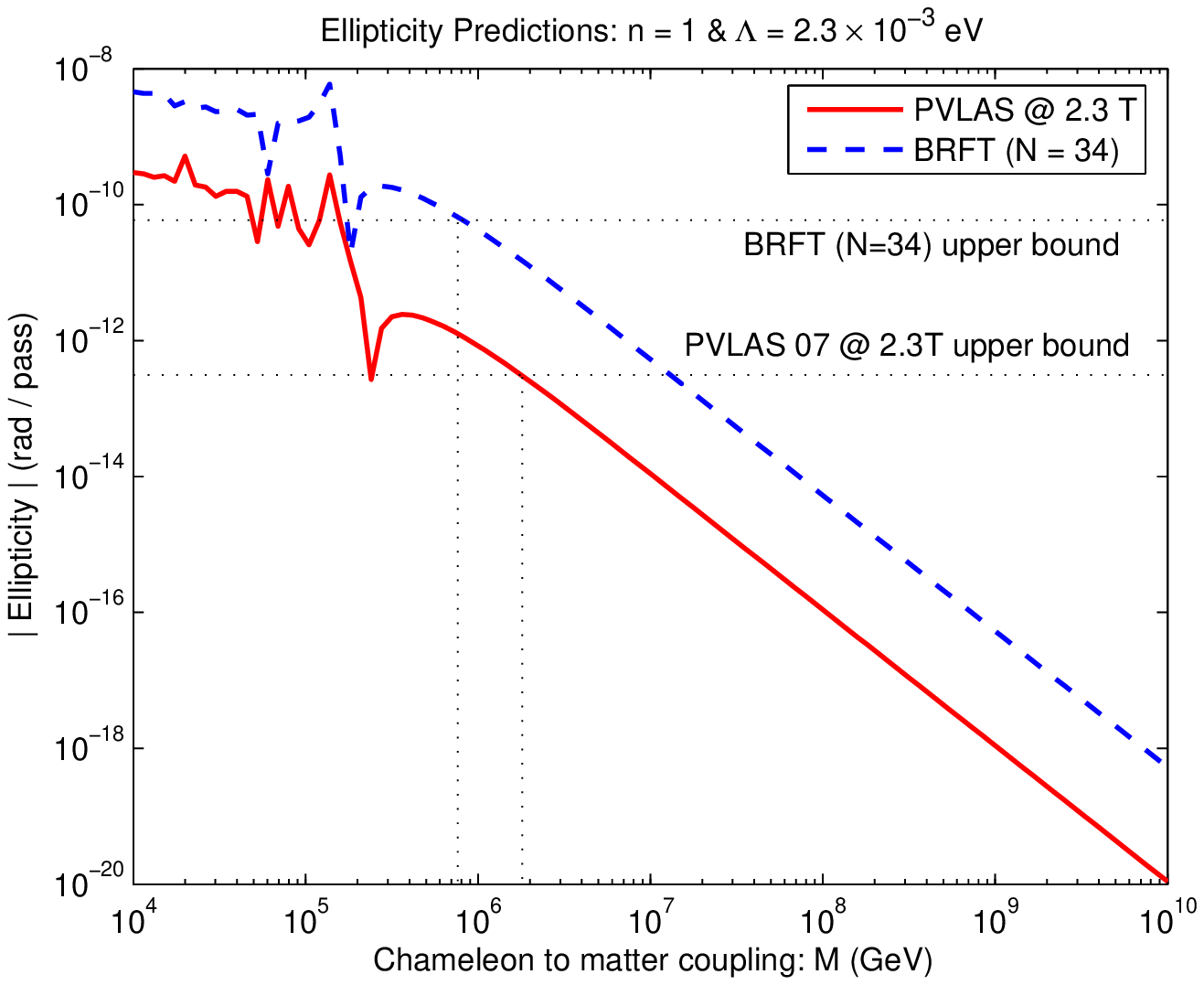}
\caption[]{Predictions for rotation (left) and ellipticity (right) in the chameleon model as a function of $M$ for $\Lambda = 2.3 \times 10^{-3}\,{\rm eV}$ 
and $n=1$. Predictions for the $B=2.3$~T PVLAS ($L=100$~cm, $d=270$~cm, $\omega=1.2$~eV, $\rho_{\rm gas} = 2\times 10^{-14}{\rm g cm}^{-3}$ 
and $\varphi=\pi/4$) and BRFT ($B = 2$~T, $L=800$~cm, $d=345$~cm, $\omega = 2.41$~eV, $\rho_{\rm gas} \approx 10^{-14}{\rm g cm}^{-3}$ and $\varphi=\pi/4$) 
set-ups are shown.  The thin-dotted lines show the $95\%$ confidence upper bounds on both the rotation and the ellipticity. }
\label{fig:BRFT}
\end{figure}  
For completeness, we show in Figure \ref{fig:BRFT} the predicted
rotation and ellipticity signals for PVLAS ($B=2.3$~T) and BRFT (for two 
different number of passes $N$) again with $n=1$ and $\Lambda = 2.3 \times
10^{-3}\eV$.  Notice that, although chameleon models predict a higher rotation 
and ellipticity within the BRFT set up, we see that BRFT should only be able 
to detect, or rule out, such chameleon theories with 
$M \lesssim 8 \times 10^{5}\GeV$. This is worse than that provided by PVLAS. 
Even though the BRFT and PVLAS set-ups use vacuums of similar quality, 
the smaller number of passes in the BRFT experiment result in it placing 
a much weaker bound on the ellipticity per pass than the PVLAS bound.

\subsection{Dependence of predictions on $n$}
\begin{figure}[htb!]
\includegraphics*[width=8.5cm]{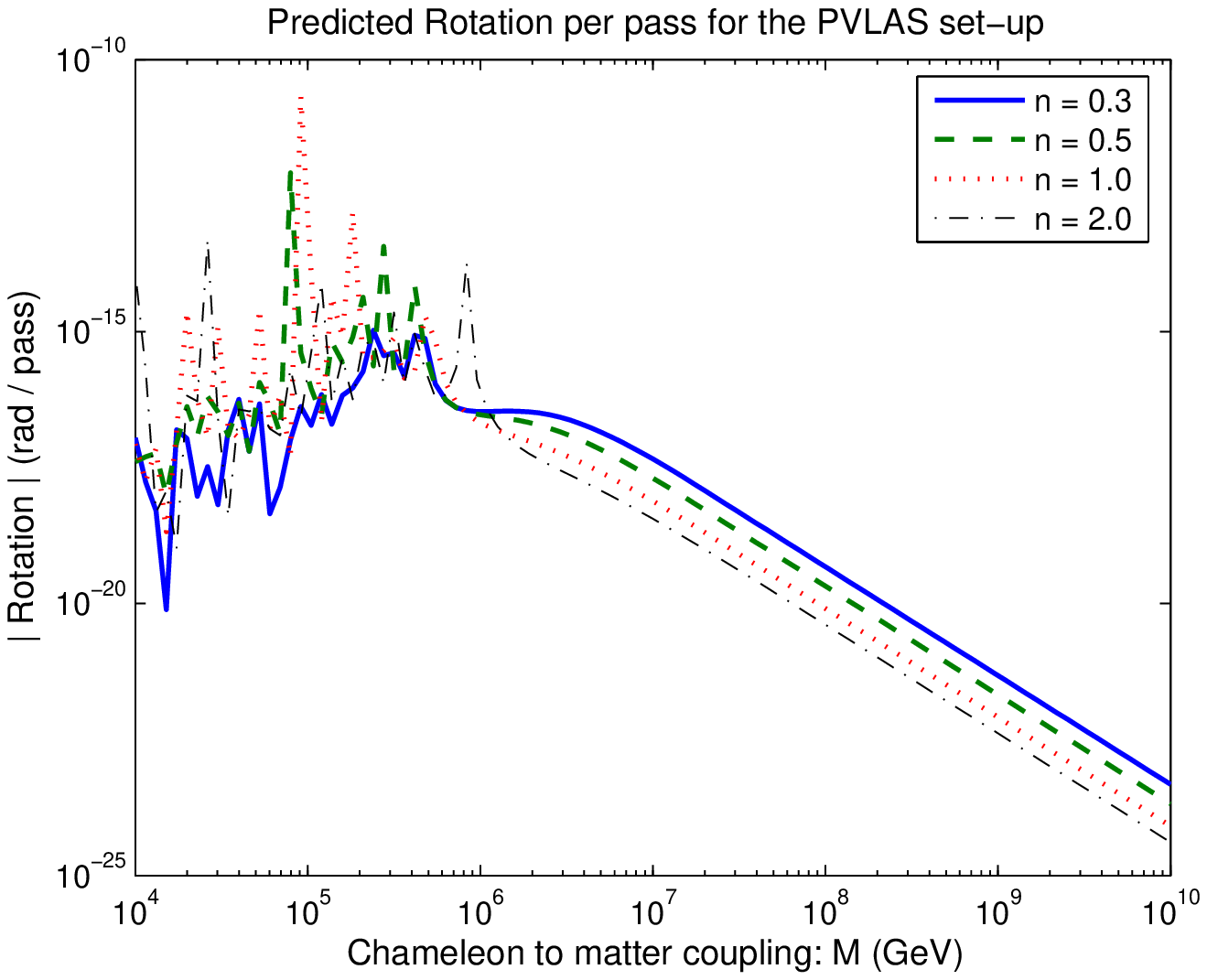}
\includegraphics*[width=8.5cm]{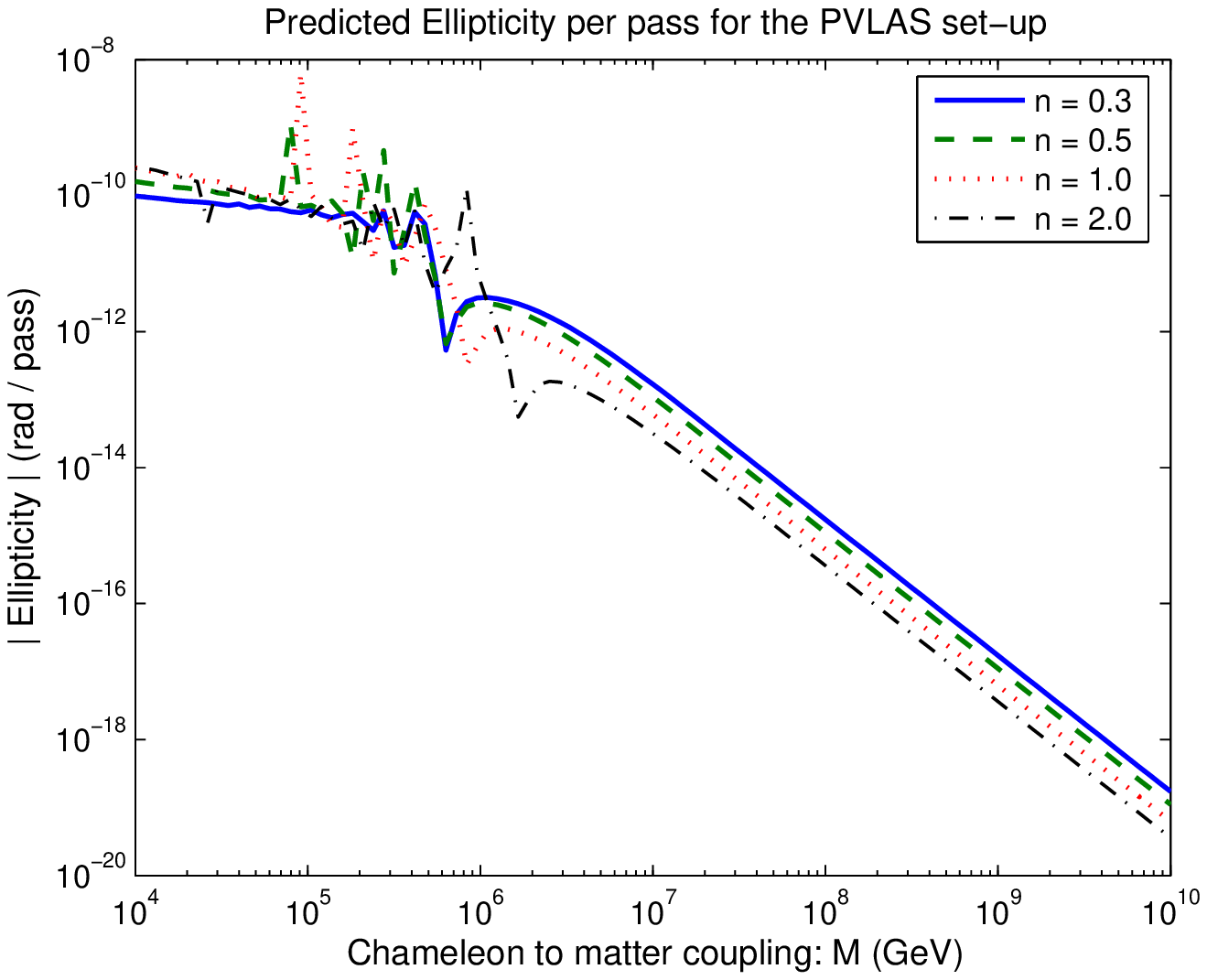}
\caption[]{Predictions for rotation (left) and ellipticity (right) in the chameleon model as a function of $M$ for $\Lambda = 2.3 \times 10^{-3}\,{\rm eV}$ and different values of $n$. The PVLAS set-up is used ($L=100$~cm, $d=270$~cm, $\omega=1.2$eV, $B=5$~T, $\rho_{\rm gas} = 2\times 10^{-14}{\rm g cm}^{-3}$ and $\varphi=\pi/4$).}
\label{fig:ncomp}
\end{figure}
In Figures \ref{fig:rotellip}-\ref{fig:PVLASBMV} we took $n=1$.
 Theories with different values of $n$ do, however, lead to different predictions.
  Figure \ref{fig:ncomp} shows how the ellipticity and rotation predicted for the PVLAS set-up depend on $n$.
   We found above that if $X = m^2_{\phi}(d+L/2)/2\omega \ll 1$ and $\tan(\Delta_{r}/2) \sim {\mathcal{ O}}(1)$ then:
$$
\psi \approx - \frac{B^2 L^2 N}{16 M^2 \tan(\Delta_{r}(n)/2)}(1+{\cal O}(X)) \approx N\sin(\Delta) \Delta\varphi,
$$
which depends on $n$ only through $\Delta_{r}(n) = \pi n/(n+2)$.
Thus for values of $M$ such that $X \ll 1$, which for PVLAS
roughly corresponds to $M \gg 10^{6} \GeV$, both the ellipticity
and rotation are therefore only very slightly dependent on $n$ and
scale as $B^2/M^2$.  Different ${\cal O}(1)$ values of $n$ do not,
therefore, alter the magnitude or $M$ dependence of either $\psi$
or $\Delta \varphi$ for $M \gg 10^{6}\GeV$.  If $M \lesssim
\mathcal{O}(10^{6}) \GeV$ however, then $X \gtrsim \mathcal{O}(1)$
then both $\psi$ and $\Delta \varphi$ depend strongly on
$m_{\phi}$ which in turn depends strongly on $\rho_{\rm gas}$, $B$ and
$n$ for such values of $M$. We can clearly see this transition
from strong to weak $n$ dependence in Figure \ref{fig:ncomp}.

\subsection{Has a chameleon field already been detected?}
The PVLAS 2007 results  found no evidence for any ellipticity with
$B=2.3$~T. However, at $B=5.5$~T a non-zero ellipticity was
detected: $\psi = (9.7 \pm 1.3) \times 10^{-8}$ with $45000$
passes \cite{PVLAS07}.  This is equivalent to an ellipticity per
pass of: $(2.2 \pm 0.3) \times 10^{-12}$.  If, as is the case for
standard ALPs, $\psi \propto B^2$ then such an ellipticity at
$5.5$~T implies that at $2.3$~T, PVLAS should have found $\psi =
(1.7 \pm 0.2) 10^{-8}$ which is, in fact, ruled out with 99\%
confidence.  A $B^2$ scaling of $\psi$ therefore implies that the
signal detected with $B=5.5$~T must be of instrumental origin
\cite{PVLAS07}.

In chameleon theories, however, $m^2_{\phi}$ can depend on $B$ and
so a $B^2$ scaling of the ellipticity is not assured. We noted
above that if $X = m^2_{\phi}(2d+L)/2\omega \ll 1$ then both
$\psi$ and $\Delta \phi$ scale as $B^2$.  If we wish to reconcile
the $5.5$~T detection with the $2.3$~T null result we must therefore
concentrate on ${\cal{O}}(1)$ or greater values of $X$, which for
PVLAS corresponds to $m_{\phi} \gtrsim 3 \times 10^{-4}\eV \gg 2/R
\approx 3\times 10^{-5}\eV$.  We are therefore well into the
region where $m_{\phi}$ depends strongly on both
$\rho_{\rm gas}+B^2/2$. Now $m_{\phi}$ grows with $B^2$, but it is
generally the case that $\vert \psi \vert$ decreases as $m_{\phi}$
grows, and therefore $\psi$ generally grows more slowly than $B^2$
for $X \sim \mathcal{O}(1)$. However, we need a faster than $B^2$
growth if we are to find a theory that predicts the signal seen at
$B=5.5$~T without violating the $B=2.3$~T upper bound.
Fortunately, the necessarily faster than $B^2$ growth in $\vert
\psi \vert$ does occur in chameleon field theories when:
$$
\frac{\Delta_{r}}{2}  + \frac{m^2_{\phi}(B) 2d + L}{4\omega} \approx m \pi,
$$
for some integer $m$.  Both $\psi$ and $\Delta \phi$ are strongly peaked about values
 of $m_{\phi}$ that satisfy the above equation.  In these cases all the effects that usually result in the chameleon and photon fields becoming decoherent cancel each other out, which results in a greatly amplified signal. For these values of $m_{\phi}$ the experiment could be thought of as being resonant.  The positions of these resonances are highly dependent on the set-up of the experiment. If $m_{\phi}(B=5.5\,{\rm T})$ for $B= 5.5$~T just so happens lies close to such a resonance but $m_{\phi}(B=2.3\,{\rm T})$ does not, then we would have $\vert \psi(B = 5.5\,{\rm T}) \vert \gg (5.5/2.3)^2 \vert \psi(B = 2.3\,{\rm T}) \vert$.   Additionally, since $\psi$ changes sign as one passes through such a resonant point, we can ensure that $\psi(B=5.5\,{\rm T}) > 0$. In the context of standard ALPs, $\psi > 0$ implies the existence of a pseudo-scalar, in the chameleon model, however, this is not the case and $\psi$ can be both positive and negative depending on the value of $m_{\phi}$.

We find that for $\Lambda = 2.3 \times 10^{-3}\eV$, there exist
values of $M$ for which $\psi(5.5\,{\rm T}) = 9.7 \pm 1.3 \times
10^{-8}$ and $\vert \psi(2.3\,{\rm T}) \vert < 1.4 \times 10^{-8}$
for $n \gtrsim 2$. In all cases $M \sim O(10^{6}\GeV)$ and $m_\phi
\sim O(10^{-3}\,{\rm eV})$ with larger values of $n$ corresponding
to both larger values of $M$ and smaller values of $\vert
\psi(2.3\,{\rm T}) \vert$.  For example: If $n=2$ then $M = (1.13
\pm 0.03) \times 10^{6}\GeV$ gives $\psi(5.5\,{\rm T}) = 9.7 \pm
1.3 \times 10^{-8}$, $\psi(2.3\,{\rm T}) = -(1.32 \pm 0.05) \times
10^{-8}$, $\Delta \varphi(5.5\,{\rm T}) = -(1.4 \pm 0.2) \times
10^{-12}$ and $\Delta \varphi(2.3\,{\rm T}) = -(1.6 \pm 0.1)
\times 10^{-13}$. If this is the case then the BMV experiment
would measure $\psi(BMV) \approx -7.3 \times 10^{-8}$. If however
$n=3$ then $M = (1.58 \pm 0.04) \times 10^{6}\GeV$ is required and
$\vert \psi(2.3\,{\rm T})\vert < 4 \times 10^{-9}$; BMV would
measure $\psi(BMV) \approx -3.5 \times 10^{-7}$.   We have checked that these models would are not ruled out by the BRFT experiment.  In all cases, the ellipticity and rotation predicted by these models in the context of that set-up is well below the BRFT upper bounds. This is due in part to the long length of the BRFT interaction region ($8$~m) which results in its sensitivity being peaked for particles with mass smaller than $1$~meV. In contrast, all of the chameleon models that reproduce the signal observed by PVLAS have $m > 1$~meV i.e. outside of the region where BRFT works best..

In the chameleon model then $\psi(5.5\,{\rm T}) = 9.7 \pm 1.3
\times 10^{-8}$ is not necessarily excluded by $\vert
\psi(2.3\,{\rm T}) \vert < 1.4 \times 10^{-8}$, however it does
require that chameleon mass for $B=5.5\,{\rm T}$ lies very close
to a resonant point.  Since the position of these resonances is
highly dependent on the set-up of the experiment this would be a
remarkable coincidence.  This issue will certainly be settled by
the upcoming BMV experiment.

\section{Discussion and Conclusions}\label{sec:con}
In this paper we have studied how chameleon fields that couple to
the electromagnetic sector alter the propagation of photons in a
vacuum.  Just as other axion like particles (ALPs) do, such
chameleon fields would induce both dichroism (rotation) and
birefringence (ellipticity) in a photon beam travelling through a
magnetic field.   Both of these effects could be detected by
laboratory searches for ALPs such as PVLAS \cite{PVLAS}, Q$\&$A
\cite{QA} and BMV \cite{BMV,BMV1}. The mass of a chameleon field depends
on its environment; specifically it is larger in backgrounds where
the ambient matter density is high than it is in those where the
background density is low. The mass and coupling strength of
standard (i.e. non-chameleonic) ALPs are strongly constrained by
limits on solar axion production.  As was pointed out in
\cite{chamPVLAS}, however, the density dependence of the chameleon's
mass implies that $m_{\phi}$ in the Sun is generally much larger
than the value of $m_{\phi}$ in the laboratory vacuum.  Solar axion
production therefore represents a far less stringent constraint on
chameleon ALPs than it does on standard ALPs.  This opens to the
door to the prospect that chameleons fields, if they exist, may well
be detected first by ongoing and upcoming laboratory axion searches.

The original motivation for this work was the reported detection of
light polarization rotation in the vacuum in the presence of a
magnetic field by the PVLAS experiment \cite{PVLAS}.  If one wishes
to explain this detection by the presence of a standard ALPs then it
must have mass $m_{ALP} \approx 1\,{\rm meV}$ and photon coupling $M
\approx 10^{6} \GeV$.  A standard ALP with these properties is,
however, strongly ruled out by bounds on solar axion production
\cite{chamPVLAS}, and if the ALP is a scalar field then it is
additionally ruled out by short-range laboratory tests of gravity.
However, as was shown in \cite{chamPVLAS}, a chameleon field with
these properties is \emph{not} ruled out.  If chameleon fields
behaved the same as standard ALPs they could, at first sight,
explain the PVLAS detection.

In this paper we have shown that, within the confines of
experiments like PVLAS, chameleon fields and standard ALPs behave
very differently. As a photon beam moves through the experiment it
generates ALPs. In the absence of a chameleon mechanism these ALPs
pass through the mirrors at either end of the Fabry-Perot cavity
and escape the experiment. Chameleon fields, however, do not
escape.  As a direct result of their chameleonic properties, the chameleon mass
$m_{\phi}$ in the mirrors is many orders of magnitude larger than
$m_{\phi}$ in the cavity. The mirrors therefore act as a potential well
for the chameleon particles and, as was shown
above, these particles do not have energy to propagate through the
mirrors. Not only then do the  photons reflect off the mirrors,
but so also do the chameleon particles.  Furthermore, whereas the
photons reflect off the surface of the mirror, a beam of chameleon
particles with energy $\omega$ was found to reflect at a distance
${\cal O}(1/\omega)$ away from the surface of the mirror.  A
coherent beam of photons and chameleons incident on the mirror is
therefore decoherent after reflection.  The presence of these two
effects means that the standard expressions for the rotation
and ellipticity no longer hold. In Section \ref{sec:prop} we
therefore derived new expressions for the dichroism and
birefringence induced in a photon beam by the presence of a
chameleon field.  The new expressions for the rotation and
ellipticity are given respectively by Eqs. (\ref{rotpredicteqn})
and (\ref{ellippredicteqn}).

The combination of both the reflection and decoherence of the
chameleon particles was seen to lead to the magnitude of the
predicted ellipticity always being much larger than that of the
predicted rotation. Specifically, if the photon beam makes $N$
passes through the Fabry-Perot cavity, then the ratio of the
rotation to the ellipticity is ${\cal O}(1/N) \ll 1$.  The large
ellipticity to rotation ratio is a generic feature of chameleon
theories, and so experimental searches for birefringence are
better placed to detect / rule out the existence of chameleon
particles than measurements of rotation.

In Section \ref{sec:predictions} we used our expressions for the
rotation and ellipticity to make predictions for the PVLAS
\cite{PVLAS}, Q$\&$A \cite{QA} and upcoming BMV \cite{BMV,BMV1}
experiments.  The magnitude of the potentially detectable signal was
found to depend heavily on the density of the laboratory vacuum. The
relatively high density vacuum used in the Q$\&$A experiment was
seen to result in a much smaller predicted rotation and ellipticity
than that for the PVLAS and BMV set-ups. If we assume that the
non-zero ellipticity detected by PVLAS with $B=5.5~{\rm T}$ is
instrumental then the most recent PVLAS results constrain $M \gtrsim
2 \times 10^{6}\GeV$ for $n=1$ and $\Lambda = 2.3 \times
10^{-3}\eV$, with a similar bound for other ${\cal O}(1)$ values of
$n$.  BMV will be able to detect or rule out the presence of
chameleon fields with $\Lambda = 2.3 \times 10^{-3}\eV$ and $M
\lesssim 3 \times 10^{7}\GeV$. On the other hand, if the $B=5.5$~T
ellipticity signal is physical, then chameleon theories can provide
an explanation with a mass $m_\phi\approx 10^{-3}$~eV and an inverse
coupling $M\approx 10^6$~GeV. In this case, a large ellipticity
should be observed by the BMV experiment, providing a decisive tests
of chameleon theories.

In conclusion, chameleon theories can be tested with light
propagating in vacuum through a magnetic field as long as
the coupling to the electromagnetic sector is large enough
($M \ll M_{\rm Pl}$). Chameleon theories are then an additional interesting
class of models to be probed by experiments of PVLAS-type. We found that
the predictions of chameleon theories are substantially different from those of
standard ALPs and as such could potentially reconcile astrophysical with local
tests. These experiments provide additional constraints on the parameter of the theory, which are
complementary to gravitational and/or Casimir-experiments.

\appendix

\section{Generalized Calculation of Propagation of Light and the Chameleon in a Cavity}\label{appA}
In Section \ref{sec:prop} we calculated how light and the chameleon field propagate in a cavity under the assumptions there is no gap between the mirrors and the ends of the interaction region and if the photon and chameleon field reflect in the same manner.  In general neither of these two assumptions hold.  The calculation performed in Section \ref{sec:prop} is very direct and elegant however we found that it becomes significantly more complicated when the assumptions are dropped. In the limit where both of the aforementioned assumptions hold, the results found using this alternative approach are entirely equivalent to those found in Section \ref{sec:prop}. In this appendix our method is based on that used in \cite{axphoprop}.

Waves in the photon field, $\mathbf{A}$, and in the chameleon field, $\phi$, obey:
\begin{eqnarray}
\square \mathbf{A} &=& \frac{\nabla \phi \times \mathbf{B}}{M}, \\
\square \phi - m^2_{\phi} \phi &=& \frac{\mathbf{B} \cdot (\nabla \times \mathbf{A})}{M}.
\end{eqnarray}
We take $\mathbf{B} = B \mathbf{e}_{x}$, $\mathbf{A} = a_{\parallel}\mathbf{e}_{x} + a_{\perp}\mathbf{e}_{y}$.  The waves then travel in the $z$-direction.   We take the interaction region, where $B \neq 0$, to have length $L$.  There is a distance $d$ between the ends of the interaction region and the mirrors.

For right moving waves $\propto \exp(ikz)$, in the interaction region, we have:
$$
\partial_{t}^2 a_{\parallel} = -k^2 a_{\parallel},
$$
and
\begin{equation}
-\partial_{t}^2 \mathbf{v} = U_{r}\mathbf{v},
\label{righteqn}
\end{equation}
where
\begin{eqnarray*}
\mathbf{v} &=& \left(\begin{array}{c} a_{\perp} \\ \chi \end{array}\right), \\
U_{r} &=& \left( \begin{array}{cc} k^2 & -kB/M \\ -kB/M & k^2+m^2 \end{array}\right),
\end{eqnarray*}
and where $\chi = i\phi$.  The eigenvectors of $U_{r}$ are:
$$
\left(\begin{array}{c} \cos \theta \\ \sin \theta \end{array} \right), \qquad \left(\begin{array}{c} -\sin \theta \\ \cos \theta \end{array} \right),
$$
with eigenvalues $\omega_{-}^2$ and $\omega_{+}^2$ respectively where:
\begin{equation}
\omega_{\pm}^2 = k^2 + m^2 \frac{\cos 2\theta \pm 1}{2 \cos 2\theta}.
\end{equation}
As in Section \ref{sec:prop} we have defined:
$$
\tan 2\theta = \frac{2B k}{Mm^2}.
$$
Eq. (\ref{righteqn}) can then be written as:
$$
-\partial_{t}^2 \mathbf{v} = Q^{T}_{+} \Omega Q_{+} \mathbf{v},
$$
where
$$
Q_{+} = \left( \begin{array}{cc} \cos \theta & -\sin \theta \\ \sin \theta & \cos \theta \end{array} \right),
$$
and $\Omega = \mathrm{diag}(\omega_{-}^2,\,\omega_{+}^2)$.  A similar equation holds for the left-moving modes:
$$
-\partial_{t}^2 \mathbf{v} = Q^{T}_{-} \Omega Q_{-} \mathbf{v},
$$
where
$$
Q_{-} = \left( \begin{array}{cc} -\cos \theta & \sin \theta \\ \sin \theta & \cos \theta \end{array} \right).
$$
In the interaction region then the evolution of the right moving modes is given by:
$$
\mathbf{v}(t) = e^{-i\omega t}V_{+}(t) \mathbf{v}(0),
$$
and the evolution of the left moving modes by:
$$
\mathbf{v}(t) = e^{-i\omega t}V_{-}(t) \mathbf{v}(0),
$$
where $V_{\pm} = Q^{T}_{\pm} \mathrm{diag}\left(e^{-i(\omega_{-}-\omega)t},\,e^{-i(\omega_{+}-\omega)t}\right) Q_{\pm}$. For $t=L$ we have:
\begin{equation}
V_{\pm}(L) = \left(\begin{array}{cc}  c_{+} &  \pm b \\ \pm b & c_{-}\end{array}\right),
\end{equation}
where:
\begin{eqnarray}
c_{\pm} &=& \cos^2 \theta e^{-i(\omega_{\mp}-\omega)L} + \sin^2 \theta e^{-i(\omega_{\pm}-\omega)L}, \\
b &=& \left(e^{-i(\omega_{-}-\omega)L} - e^{-i(\omega_{+}-\omega)L}\right) \frac{\sin 2\theta}{2}.
\end{eqnarray}
Thus a wave that enters the magnetic field region at $z=0$ as $\mathbf{v}_{0}$ will, upon exiting, the field region at $z=L$ have evolved into:
$$
\mathbf{v}(L) = e^{-i\omega L} V_{+}(L) \mathbf{v}_{0}.
$$
The wave then exits the magnetic field, travels a distance $d$ to the mirror, is reflected, travels a further distance $d$ and re-enters the magnetic field at $z=L$.  Outside the interaction region the chameleon field travels more slowly than the photon, and furthermore it may reflect at a different point.  In general then the chameleon will return to $z=L$ having been shifted by a phase relative to the photon field.  We define this phase to be $\Delta$.   The assumption made in Section \ref{sec:prop} are equivalent to taking $\Delta = 0$.  We calculate this $\Delta$ in Appendix \ref{appB}.

When the fields re-enters the field region, this time moving leftwards, they are therefore (up to an overall irrelevant phase factor) in a state:
$$
\mathbf{v} = e^{-i\omega L} R V_{+}(L) \mathbf{v}_{0},
$$
where $R = \mathrm{diag}(1,\,-\exp(-i\Delta))$.  The $-$ sign is in the last component of $R$ due to the fact that chameleons are scalar fields with positive parity, whereas the photon fields are vectors with negative parity. The fields then travels through the interaction region, this time moving leftwards, exit the region and reflect back again, so that by the time they have returned to $z=0$ moving rightwards they are in a state:
$$
\mathbf{v}_{round} = e^{-2i\omega L} \left(R V_{-} R V_{+}\right) \mathbf{v}_{0}.
$$
If $\Delta$ is taken to have an imaginary part then this analysis also allows us to account for any scalar fields that escape the cavity. For instance if, as is often taken to be case when the ALP is not a chameleon, the scalar field does \emph{not} reflect then $\Delta = -i\infty$, and $R = \mathrm{diag}(1,\,0)$. We define:
\begin{equation}
F = RV_{-} RV_{+} = \left(\begin{array}{cc} \alpha  & \gamma \\ e^{-i\Delta} \gamma & e^{-i\Delta}\beta\end{array}\right),
\end{equation}
where
$$
\alpha = c_{+}^2 + e^{-i\Delta}b^2, \quad \beta = c_{-}^2 e^{-i\Delta} + b^2, \quad \gamma = b\left(c_{+} + e^{-i\Delta}c_{-}\right).
$$
Then after $N$ passes through the interaction region, the fields will be in a state:
$$
\mathbf{v}_{N} = e^{-2i\omega NL} F^{N} \mathbf{v}_{0}.
$$
We define
$$
F^{N} \mathbf{v}_{0} = \left( \begin{array}{c} A_{N} \\ Z_{N} \end{array} \right),
$$
and find that:
\begin{eqnarray}
A_{N+1} = \alpha A_{N} + \gamma Z_{N}, \\
Z_{N+1} = e^{-i\Delta} \gamma A_{N} + e^{-i\Delta}\beta Z_{N},
\end{eqnarray}
and so:
\begin{eqnarray}
A_{N+1} = \left(\alpha + \beta e^{-i\Delta}\right)A_{N} + e^{-i\Delta}\left[\gamma^2 - \alpha \beta\right] A_{N-1}.
\end{eqnarray}
We take initial conditions $A_{0} = \sin \varphi$ and $Z_{0} = 0$ which implies that $A_{1} = \alpha \sin \varphi$.  We then find the solution:
\begin{equation}
\frac{A_{N}}{\sin \varphi} = \left(\frac{\alpha - \mu_{-}}{\mu_{+}-\mu_{-}} \mu_{+}^{N}  - \frac{\alpha - \mu_{+}}{\mu_{+}-\mu_{-}} \mu_{-}^{N}\right),
\end{equation}
where
\begin{eqnarray}
\mu_{+} = \alpha + e^{-i\Delta} \delta \mu, \\
\mu_{-} = e^{-i\Delta}\left(\beta  - \delta \mu\right),
\end{eqnarray}
where
$$
e^{-i\Delta}\delta \mu = \frac{\alpha - \beta e^{-i\Delta}}{2} \left( (1+e^{-i\Delta} \nu^2)^{1/2} -1 \right),
$$
and
$$
\nu = \frac{2\gamma}{\left(\alpha - \beta e^{-i\Delta}\right)} = \frac{2b}{c_{+} - e^{-i\Delta} c_{-}}.
$$
If $\Delta = 0$ then:
$$
\mu_{\pm} = e^{-i(\omega_{\mp}-\omega)L}.
$$
We take $\theta \ll 1$ and so:
$$
e^{-i\Delta}\delta \mu \approx  \frac{\gamma^2
e^{-i\Delta}}{\alpha - \beta e^{-i\Delta}} =
\frac{e^{-i\Delta}b^2(c_{+}+e^{-i\Delta}c_{-})}{c_{+}-c_{-}e^{-i\Delta}}
\propto \theta^2 \ll 1.
$$
We define $\delta_{\pm}/2\omega = \omega_{\pm}-\omega$ and $\delta = \delta_{+}-\delta_{-}$. For small $\theta$, $\delta_{-} = {\cal O}(\theta^2)$.

After $N$ round trips, a photon which was initial in a state $(\cos \varphi, \sin \varphi)^{T}$ has evolved into:
$$
\left( \begin{array}{cc} a_{\parallel}(N) \\ a_{\perp}(N) \end{array} \right) = \left( \begin{array}{cc} \eta(N) \cos (\varphi+\Delta \varphi(N)) \\ \eta(N) \sin (\varphi + \Delta \varphi(N))e^{i \rho(N)} \end{array} \right) = \left( \begin{array}{cc} \cos \theta \\ \frac{\tan(\varphi + \Delta \varphi(N))}{\tan \varphi} e^{i \rho(N)} \sin \theta \end{array} \right),
$$
where
\begin{equation}
\frac{\tan (\varphi + \Delta \varphi(N))e^{i\rho(N)}}{\tan \varphi} = \frac{A_{N}}{\sin \theta}.
\end{equation}
Since the photon is trapped in the cavity, what one actually measures is a superposition of states which have each completed a different number of passes.  One therefore measures:
$$
\mathbf{a}_{measured} = \frac{1}{N+1}\sum_{k=0}^{N} \mathbf{a}(k)
=  \left( \begin{array}{cc} \cos \theta \\ \frac{\tan(\varphi +
\Delta \varphi_m)}{\tan \varphi} e^{i \rho_m} \sin \theta
\end{array} \right),
$$
where
$$\frac{\tan (\varphi + \Delta \varphi_m)e^{i\rho_m}}{\tan \varphi} = \frac{1}{N+1} \sum_{k=0}^{N} \frac{A_{k}}{\sin \theta}.$$
We calculate $\Delta \phi_{m}$ and $\rho_{m}$ to ${\cal O}(\theta^2)$. To this order
\begin{eqnarray}
\mu_{-} \approx e^{-2i \Delta } e^{-i \delta_{+} L / \omega} \left( 1 + 2\theta^2  \left[\frac{(1-e^{-i\Delta})(1-e^{-i\delta L / 2\omega}) }{1-e^{-i\Delta-i\delta L/2\omega}}\right]\right), \\
\mu_{+} = e^{-i\delta_{-}L/\omega} \left( 1 - 2\theta^2 \left[\frac{(1-e^{-i\Delta})(1-e^{-i\delta L / 2\omega}) }{1-e^{-i\Delta-i\delta L/2\omega}}\right]\right).
\end{eqnarray}
We define
$$
H_{\Delta}(x) = \left(\frac{\sin(x/2)}{\sin(\Delta/2 + x/2)}\right)^2.
$$
Assuming that $\theta^2 H_{\Delta}(\delta L / 2\omega) \ll 1$, we perform the sum over the $A_{k}$ and find that
 we find:
\begin{eqnarray}
\frac{\tan (\varphi + \Delta \varphi_m)e^{i\rho_m}}{\tan \varphi} =&& 1-\theta^2 H_{\Delta}\left(\frac{\delta L}{2\omega}\right)  - iN \left[\frac{\delta_{-}L}{2\omega} + \theta^2 G_{\Delta}\left(\frac{\delta L}{2\omega}\right)\right] \\ &+& \theta^2 e^{-iN \Delta - \frac{iN \delta L}{2\omega}} H_{\Delta}\left(\frac{\delta L}{2\omega}\right)\delta_{N}\left(\Delta + \frac{\delta L}{2\omega}\right) \nonumber,
\end{eqnarray}
where we have defined
$$
G_{\Delta}(x) = \frac{2 \sin(\Delta/2)\sin(x/2)}{\sin(\Delta/2 + x/2)},
$$
and
$$
\delta_{N}(x) = \frac{\sin((N+1)x}{(N+1)\sin x}.
$$
If $\Delta = 0$ then $G_{\Delta} = 0$ and $H_{\Delta} = 1$. The rotation is given by $\Delta \phi_{m}$ and the ellipticity by $\psi = -\rho_{m}\sin 2\varphi/2$. For general real $\Delta$ we therefore find:
\begin{eqnarray*}
\frac{\Delta \varphi}{\sin 2\varphi} = &&-\left(\frac{B \omega}{M m_{\phi}^2}\right)^2 H_{\Delta}\left(\frac{m^2_{\phi} L}{2\omega}\right)\left\lbrace\frac{1}{2} + \left[\sin^2\left(\frac{N\Delta}{2} + \frac{N m^2_{\phi} L}{4\omega}\right)\right.\right.\\&&-\left.\left.\frac{1}{2}\right]\delta_{N}\left(\Delta+\frac{m^2_{\phi} L}{2\omega}\right)\right\rbrace,
\end{eqnarray*}
and the ellipticity is
\begin{eqnarray*}
\frac{\psi}{\sin 2 \varphi} = &&= -\frac{1}{2}\left(\frac{B \omega}{M m_{\phi}^2}\right)^2 \left\lbrace \frac{Nm^2_{\phi} L}{2\omega} - N G_{\Delta}\left(\frac{m^2_{\phi} L}{2\omega}\right)\right. \\ &&- \left. \sin\left(N\Delta+\frac{Nm^2_{\phi} L}{2\omega}\right) H_{\Delta}\left(\frac{m^2_{\phi} L}{2\omega}\right)\delta_{N}\left(\Delta+\frac{m^2_{\phi} L}{2\omega}\right)\right\rbrace.
\end{eqnarray*}
As a consistency check we can also consider the case where the scalar fields escape after every pass. This is given by  taking $\exp(-i\Delta) \rightarrow 0$, and in this limit $H_{\Delta} \rightarrow 0$ and $G_{\Delta}(x) \rightarrow \sin(x)  2i\sin^2(x/2)$ and so we recover the standard formulae:
$$
\frac{\Delta \varphi_{escape}}{\sin 2 \varphi} = -\left(\frac{B \omega}{M m_{\phi}^2}\right)^2 \sin^2\left(\frac{m^2_{\phi} L}{4\omega}\right),
$$
and
$$
\frac{\psi_{escape}}{\sin 2\varphi} = -\frac{1}{2}\left(\frac{B \omega}{M m_{\phi}^2}\right)^2 \left[ \frac{Nm^2_{\phi} L}{2\omega} - N\sin\left(\frac{m_{\phi}^2 L}{2\omega}\right)\right].
$$
In Appendix \ref{appB} we consider the propagation and reflection of the chameleon field outside the interaction region and evaluate $\Delta$.

\section{Reflection of the Chameleon Field}\label{appB}
We consider how chameleon waves with frequency $\omega$ reflect off a flat mirror that is placed at $z=z_0$, and propagate relative to a photon field outside the interaction region. We write $\phi = \phi_{0}(z) + \delta \phi(z,t)$, where $\delta \phi$ is a small perturbation with frequency $\omega$, and $\phi_{0}$ is the background value of $\phi$. We begin by showing that $m_{\phi} \sim 1/(z-z_{0})$ near the mirror.

\subsection{Behaviour of the chameleon mass near a mirror}
We define $\rho_{m}$ to be the density of the mirror and define
$\phi_{m}$ by:
$$
V^{\prime}(\phi_m) = -\frac{\rho_{m}}{M}.
$$
We define $m_{m} = m_{\phi}(\phi_m)$. We take the mirror to lie in the region $z < z_0$ with its surface at $z=z_0$.
Since $M \ll M_{Pl}$, $\phi_0$ must be $\approx \phi_m$ deep inside the mirror for consistency with experimental tests of gravity. Inside the mirror, $\phi_{0}$ obeys
$$
\frac{\dd^2 \phi_{0}}{\dd z^2} = V^{\prime}(\phi_{0}) + \frac{\rho_{m}}{M},
$$
and outside the mirror ($z>z_0$), where the density of matter is $\rho_{\rm gas}$, we have
$$
\frac{\dd^2 \phi_{0}}{\dd z^2} = V^{\prime}(\phi_{0}) + \frac{\rho_{\rm gas}}{M}.
$$
We define $\phi_{c}$ by:
$$
V^{\prime}(\phi_c) = -\frac{\rho_{\rm gas}}{M},
$$
and $m_c = m_{\phi}(\phi_c)$. Integrating the above equations and assuming that deep inside the mirror $\phi \rightarrow \phi_{m}$ and $\dd \phi / \dd z \rightarrow 0$ we find:
\begin{eqnarray}
\frac{1}{2}\left(\frac{\dd \phi_0}{\dd z}\right)^2 &=& V(\phi_0(z)) - V(\phi_m) + \frac{\rho_{m}}{M}(\phi_0(z)-\phi_m) \quad z < z_0, \label{inbody} \\
\frac{1}{2}\left(\frac{\dd \phi_0}{\dd z}\right)^2 &=& V(\phi_0(z)) - V(\phi_c) + \frac{\rho_{\rm gas}}{M}(\phi_0(z)-\phi_c) \quad z > z_0. \label{outbody}
\end{eqnarray}
Matching these equations at $z=z_0$ and using $\rho_{m} \gg \rho_{\rm gas}$ we find that at $z=z_{0}$:
\begin{equation}
\phi_0(z=z_0) \approx \phi_{m} - \frac{V(\phi_m)-V(\phi_{c})}{V^{\prime}(\phi_m)}.
\end{equation}
>From Eq. (\ref{outbody}), we see that outside the mirror, in the region where $\vert V^{\prime}(\phi_c)/V^{\prime} (\phi_0) \vert \ll 1$, we have:
$$
\frac{1}{2} \left(\frac{\dd \phi_{0}}{\dd z}\right)^2 = V(\phi_{0}) - V(\phi_c).
$$
We now normalize the potential $V$ so that as $\rho_{\rm gas} \rightarrow 0$, $V(\phi_c(\rho_{\rm gas})) \rightarrow 0$, i.e. we neglect any constant term in $V$.   $\vert V^{\prime}(\phi_c)/V^{\prime} (\phi_0) \vert \ll 1$ then implies than $V(\phi_{0}) \gg V(\phi_{c})$ and so:
\begin{equation}
a_{\phi}  \frac{d (1/m_0)}{d z}  = 1,
\end{equation}
where
$$
a_{\phi} = \frac{\sqrt{2}(V^{\prime \prime}(\phi_{0}))^{3/2}}{(-V^{\prime \prime \prime}(\phi_0)) (V(\phi_{0}))^{1/2}} > 0.
$$
If $V \propto \phi^{-n}$ then $a_{\phi} = \sqrt{2n(n+1)/(n+2)^2}$.   We define $a_{\phi}(z_0) = a_{s}$ and use the shorthand $a_{\phi}(\phi(z)) = a_{\phi}(z)$, we then have:
\begin{equation}
\frac{1}{m_0} = \frac{1}{m_s} + \frac{z-z_0}{a_{\phi}(z)} + \frac{1}{a_{\phi}(z)} \int_{a_s}^{a_{\phi}(z)}\frac{1}{m_0}\,\dd a_{\phi}.
\end{equation}
In many theories, the potential is such that $a_{\phi}$ changes only very slowly  i.e. $\vert d \ln a_{\phi} / dx \vert \ll \vert d \ln m_{0} / dx \vert$ for $\omega \lesssim  m_{0} \lesssim m_{m}$; when this is true we have:
\begin{equation}
m_{0} \approx  \frac{a_{\phi}(z)}{z - z_{0} + a_{\phi}(z)/m_s}, \label{m0eqn}
\end{equation}
where $a_{\phi}(z)$ is slowly varying compared to $m_0$ and $m_{s} \equiv \sqrt{V^{\prime \prime}(\phi_{0}(z=z_0))}$ is the mass of the chameleon on the surface of the mirror.

Since we have assumed that  $\vert V^{\prime}(\phi_c)/V^{\prime} (\phi_0) \vert \ll 1$ then we must have $m_{0}(z) \gg m_{c}$ which implies $z - z_{0} \ll 1/m_{c}$.  For $1/m_{m} \lesssim 1/m_s \ll  \Delta z = z-z_0 \ll 1/m_{c}$, Eq. (\ref{m0eqn}) gives $m_{0} \propto 1/\Delta z$.

This behaviour will occur in any chameleon theory where $a_{\phi}$ varies slowly with $\phi$ compared to $m_{\phi}$. More generally, we have:
$$
m_0 \geq \frac{a_{\phi}(z)}{z - z_{0} + a_{\phi}(z)/m_s},
$$
if
$$
b_{\phi}(\phi) \equiv \frac{2V^{\prime \prime \prime \prime}V^{\prime \prime}}{3V^{\prime \prime \prime \, 2}} + \frac{V^{\prime \prime}V^{\prime}}{3V^{\prime \prime \prime}V} \leq 1,
$$
and $m_0 \leq a_{\phi} / (z- z_0 + a_\phi / m_s)$ if $b_{\phi} \geq 1$. The assumption that $a_{\phi}$ is slowly varying compared to $1/m_0$ is equivalent to $3\vert 1-b_{\phi} \vert \ll 1$ for $m_0 < m_{\phi} < m_s \lesssim m_{m}$.

\subsection{Reflection of chameleon waves}
We now consider the reflections of chameleon waves: $\delta \phi = \delta \tilde{\phi}(z) e^{-i\omega t}$.  These evolve according to:
$$
\frac{d^2 \delta \tilde{\phi}}{d z^2} \approx \left(V^{\prime \prime}(\phi_{0}) - \omega^2\right) \delta \tilde{\phi}.
$$
For most sensible choices of potential we have $m_{s} \sim O(m_m = m_{\phi}(\phi_m))$ and we assume that $m_m \gg \omega$; which must certainly be the case if $\omega \sim {\cal O}(1)\,\mathrm{eV}$ and the constraints on solar axion production are satisfied \cite{chamPVLAS}.  For  $z \lesssim z_0$ then we have $\delta \tilde{\phi} \approx C \exp(m_{eff}(z) (z-z_{0}))$ where $m_{eff}(z) \sim O(m_{\phi}(\phi_m))$ and $C$ is a constant.

The behaviour of $m_0$ for $m_0 \gg m_c \equiv m_{\phi}(\phi_c)$ is given by Eq. (\ref{m0eqn}).  In the experiments that we consider $m_c  \ll \omega$. The behaviour of $m_{0}$ for $m_{0} \ll \omega$ only effects this reflection calculation at sub-leading order and so, to a first approximation we have taken $m_0(z)$ to be given by Eq. (\ref{m0eqn}).  Defining $x = z-z_{0} + a_{\phi}/m_{s}$, we find that near the mirror we have:
$$
\frac{d^2 \delta \tilde{\phi}}{d x^2} \approx
\left(\frac{a_{\phi}^2}{x^2} - \omega^2\right) \delta
\tilde{\phi},
$$
which has solutions:
$$
\delta \tilde{\phi} = \sqrt{\omega x} \left(c_{1} J_{\alpha}(\omega x) + c_{2} N_{\alpha}(\omega x)\right),
$$
where $J_{\alpha}$ and $N_{\alpha}$ are Bessel functions and $\alpha = \frac{1}{2}\sqrt{1+4a_{\phi}^2}$.  Near $z=z_{0}$ we have $\omega x \approx a_{\phi} \omega / m_{s} \equiv \delta$.  Generally $\delta = a_{\phi} \omega / m_{s} \sim O(\omega/m_{m}) \ll 1$.  Matching at $z = z_{0}$ we find:
\begin{eqnarray}
C &=& \sqrt{\delta}(c_{1} J_{\alpha}(\delta) + c_{2} N_{\alpha}(\delta)), \\
m_{1} \delta C &=& \frac{1}{2}\sqrt{\delta}\left(
c_{1}\left(J_{\alpha}(\delta)+2\delta
J^{\prime}_{\alpha}(\delta)\right)  +
c_{2}\left(N_{\alpha}(\delta)+2\delta
N^{\prime}_{\alpha}(\delta)\right)\right),
\end{eqnarray}
where $m_{1} \sim O(m_{s})$ is given by $m_{1} = \left.d \exp(-m_{eff}(z)(z-z_0)) / dz \right\vert_{z=z_0}$.  The precise value of $m_{1}$ is generally not important.   We find then that:
$$
c_{2} = B(\delta) c_{1},
$$
where
$$
B(\delta) = \left(\frac{m_{1} \delta - \frac{1}{2} + \frac{\delta J_{\alpha}^{\prime}(\delta)}{J_{\alpha}(\delta)}}{m_{1} \delta - \frac{1}{2} + \frac{\delta Y_{\alpha}^{\prime}(\delta)}{Y_{\alpha}(\delta)}}\right) \frac{J_{\alpha}(\delta)}{Y_{\alpha}(\delta)}.
$$
In general $\delta \ll 1$ which implies that $B(\delta) \ll 1$ and so $c_{2} \ll c_{1}$.  A distance $z \gg 1/m_{\phi}(\phi_c)$ from the mirror then $\delta \phi$ is therefore given by:
\begin{equation}
\delta\tilde{\phi} \approx \sqrt{\frac{2}{\pi}} c_{1} \sin\left(\omega (z-z_0) -\left(\frac{\pi \alpha}{2} - \frac{\pi}{4}\right)\right). \label{phieqnwave}
\end{equation}
We write:
$$
\delta\phi = \phi_{0}e^{-i\omega t} \left( e^{-i k (z-d-z_0)} - e^{i k(z-d-z_0)- i \Delta_{r} + i \omega 2d}\right).
$$
A chameleon wave that leaves the interaction region, reflects off the mirror is therefore, on its return to the interaction region, shifted by a phase  $\Delta \equiv 2(k-\omega)d + \Delta_r$ relative to a photon that has made the same journey.  From Eq. (\ref{phieqnwave}) we have
\begin{equation}
\Delta_{r} = \frac{\pi}{2}\left(2\alpha - 1\right) = \frac{\pi}{2}\left(\sqrt{1+4a_{\phi}^2}-1\right).
\end{equation}
If $\Delta_{r} \ll 1$ then the additional phase-shift due to the the chameleon travelling more slowly than the speed of light (the $2(k-\omega)d$ term in $\Delta$) might be important.  Taking $k = \sqrt{\omega^2 - m_{\phi}^2(\phi_m)}$, we find:
$$
\Delta_{m} = 2(\omega - k)d \approx \frac{m_{c}^2d}{\omega}.
$$
We therefore we find:
$$
\Delta = \Delta_r + \Delta_m  \approx \frac{\pi}{2}\left(\sqrt{1+4a_{\phi}^2}-1\right) + \frac{m_{c}^2d}{\omega}.
$$
Throughout this work we have been primarily concerned with  inverse power law potentials: $V - \Lambda^4 \propto \phi^{-n}$ for $n > 0$.  Valid chameleon  theories exist with $n = -2m$, where $m$ is a positive integer.  In a more general scenario the potential might, around the point where $m_{\phi} \approx \omega$ look locally like $c_{0} + c_{1}\phi^{-n}$, for some $c_{0}$ and $c_{1}$, or any $n$.  For such theories if $n < -2$ or $n>-2/3$ then we have:
$$
\Delta_{r} = \frac{\pi n}{n+2},
$$
whereas if $-2 < n < -2/3$ we find:
$$
\Delta_{r} = -\frac{\pi (n+1)}{2(n+2)}.
$$

\section*{Acknowledgements} 

We are grateful to G. Cantatore, H. Gies, H. Mei, A. Ringwald and C. Rizzo for sharing information about the experiments and useful comments on a draft of this paper. CvdB and ACD are supported partly by PPARC
DFM is supported by the Alexander von Humboldt Foundation. DJS is supported by PPARC.

\section*{References} 

\end{document}